\documentclass[3p]{elsarticle}

\usepackage{lineno,hyperref}
\modulolinenumbers[5]

\usepackage{subcaption}
\usepackage{amssymb}
\usepackage{amsmath}

\journal{Journal of Sound and Vibration}

\begin{document}

\begin{frontmatter}

\title{Beamforming for measurements under disturbed propagation conditions using numerically calculated Green's functions.}


\author[mymainaddress]{Marius Lehmann \corref{mycorrespondingauthor}}
\ead[url]{www.ebmpapst.com}
\cortext[mycorrespondingauthor]{Corresponding author}
\ead{Marius.Lehmann@de.ebmpapst.com}

\author[mysecondaryaddress]{Daniel Ernst}
\author[mymainaddress]{Marc Schneider}
\author[mysecondaryaddress]{Carsten Spehr}
\author[mytertiaryaddress]{Markus Lummer}

\address[mymainaddress]{ebm-papst Mulfingen GmbH \& Co. KG, Bachmühle 2, D-74673 Mulfingen, Germany}
\address[mysecondaryaddress]{German Aerospace Center (DLR), Bunsenstr. 10, D-37073 Göttingen, Germany}
\address[mytertiaryaddress]{German Aerospace Center (DLR), Lilienthalplatz 7, D-38108 Braunschweig, Germany}

\begin{abstract}
Beamforming methods for sound source localization are usually based on free-field Green's functions to model the sound propagation between source and microphone. This assumption is known to be incorrect for many industrial applications and the beamforming results can suffer from this inconsistency regarding both, accuracy of source power estimation, and accuracy of source localisation. The aim of this paper is to investigate whether the use of numerically calculated Green's functions can improve the results of beamforming measurements.

The current test cases of numerical and experimental investigations consists of sources placed in a short rectangular duct. The measurement is performed outside the duct in a semi-anechoic chamber. A typical example for this kind of installation is a fan with a heat exchanger. 

The Green's functions for this test case are calculated numerically using the boundary element method. These tailored Green's functions are used to calculate the corresponding beamforming steering vectors. The weighting of the Green's functions in the steering vectors has a decisive influence on the beamforming results. A generalization of the common steering vector formulations is given based on two scalars. It is shown that arbitrary differentiable Green's functions can be used to find the correct source position or source power level by using the appropriate steering vector formulations.

Beamforming measurements are performed using a loudspeaker as a reference source at different positions in the heat exchanger duct. The measurements are evaluated in the frequency domain and by means of different validation criteria it can be shown that the results with the numerical calculated Green's function are improved compared to free field beamforming especially at low frequencies. 
\end{abstract}


\end{frontmatter}

\graphicspath{{./figures/}}


\begin{tabular}[b]{|l l l l|}
    \hline
     & \multicolumn{2}{l}{\textbf{Nomenclature}} & \\
     $a$ & complex pressure amplitude at source & $\mathcal{F}^{-1}$ & inverse Fourier transform \\
     $b$ & spatial resolution &  &  \\
     $c$ & speed of sound & $(\cdot)^*$ & complex conjugate (transpose) \\
     $d$ & diameter of ring array & $(\cdot)^H$ & hermitian transpose \\
     $f$ & frequency &  &  \\
     $g$ & Green's function & $(\cdot)_k$ & for the $k$-th sample \\
     $J$ & functional to minimize & $(\cdot)_m$ & for the $m$-th microphone \\
     $K$ & number of samples during time period & $(\cdot)_n$ & for the $n$-th microphone \\
     $k$ & sample index & $(\cdot)_{\text{max}}$ & maximum \\
     $L$ & level & $(\cdot)_{SL}$ & side lobe \\
     $M$ & number of microphones &  & \\
     $m$ & microphone index & BEM & Boundary Element Method  \\
     $Ma$ & Mach number & CAA & Computational Aeroacoustics \\
     $n_{sf}$ & number of samples in steering function & CAD & Computer Aided Design  \\
     $\mathbf{p}$ & pressure vector & CSM, $\mathbf(C)$ & Cross Spectral Matrix \\
     $p$ & pressure signal & FEM & Finite Element Method \\
     $\hat{p}$ & complex pressure amplitude & FF & free-field \\
     $r$ & distance focus point/ microphone & FIR & Finite Impulse Response (filter) \\
     $t$ & time & FM & Fast Multipole \\
     $\mathbf{w}$ & steering vector & FMCAS & Fast Multipole Code for Acoustic Shielding \\
     $\mathbf{y}$  & focus point & iFFT & inverse Fast Fourier Transform \\ 
     $x, y, z$ & cartesian coordinates  & ISM & Image Source Method \\    
     &  & MSR & main lobe to side lobe ratio \\  
     $\bar{\sigma}(t,\mathbf{y})$  & estimated source signal & ppw & points per wavelength \\ 
     $\hat{\sigma}(f,\mathbf{y})$ & estimated source auto power & SPR & source to pattern ratio \\ 
     \hline
\end{tabular}

\section{Introduction}
\label{sec:intro}
Reliable knowledge about aeroacoustic sound sources on fans plays a decisive role in the development of quiet fans. Therefore, the experimental localisation of sound sources on fan blades during rotation is an important task on the path to understanding and reducing these rotating sources. For this purpose rotating beamforming methods based on microphone array measurements are applied. The main post-processing techniques on rotating sources measured by a stationary array can be divided into two groups with regard to the implementation of rotation:

In the first group, the sound propagation from the rotating focus grid to the stationary microphones is modelled. The most important representative of this approach is the ROSI algorithm  \cite{sijtsma2001location, sijtsma2006using}. It is derived from the pressure field of a moving monopole in uniform flow. This approach is used to calculate the time dependent Green's functions from a fan-fixed focus grid to the array microphones. Any microphone array layouts can be taken into account.

In the second group, the stationary microphones in the array are virtually set into rotation. Therefore, the transfer function between the sources on the rotating blades and the virtual with the same speed rotating microphones becomes time- independent. This makes it possible to perform the beamforming process in the frequency domain. The virtual rotation itself can be performed in both, time domain and frequency domain. The virtual rotation in the time domain was first presented by Dougherty et al. \cite{dougherty2010locating} and enhanced by Herold \& Sarradj \cite{herold2015microphone}. An example of a method that works completely in the frequency domain was presented by Lowis and Joseph \cite{lowis2006focused, lowis2007duct} who investigated rotating sources in ducts. Pannert \& Maier \cite{pannert2014rotating} and Ocker \& Pannert \cite{ocker2017imaging} have adapted the method to free-space conditions. Most of the mentioned methods are limited to circular arrays. However, Jekosch \cite{jekosch2020extension} has extended the application of virtual rotation in the time domain to arbitrary array geometries. 

\begin{figure}
	\centering
	\begin{subfigure}[b]{0.3\textwidth}
		\centering
		\includegraphics[width=\textwidth]{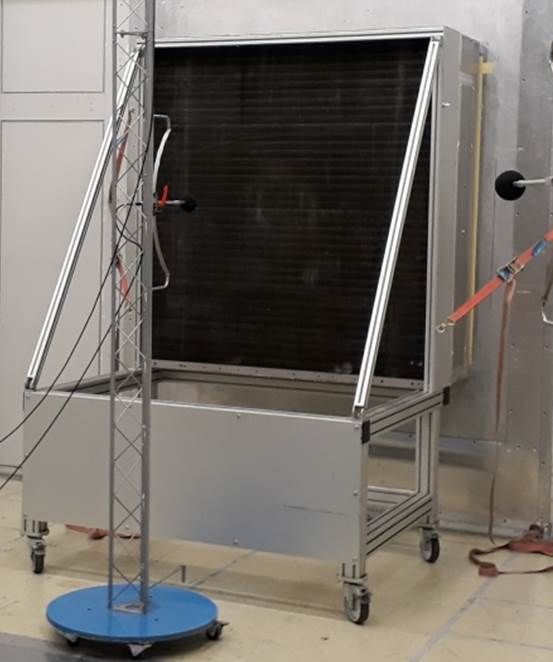}
		\caption{}
		\label{fig:WT}
	\end{subfigure}
	\begin{subfigure}[b]{0.3\textwidth}
		\centering
		\includegraphics[width=\textwidth]{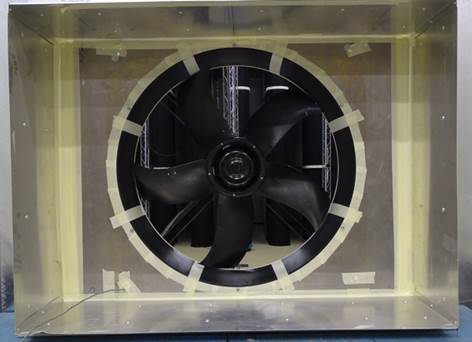}
		\caption{}
		\label{fig:box}
	\end{subfigure}
	\caption{Example of a fan mounted in a heat exchanger. (a): Suction side view of a heat exchanger. (b): Suction side view of an axial fan in a heat exchanger housing without condenser.}
	\label{fig:fan_heat_exchanger}
\end{figure}

These rotating beamforming methods can be used only for test cases where the sound propagation to the microphones takes place under free-field conditions or where both, fan and microphone array are installed under rotationally symmetric conditions. However, a typical application for axial fans is the installation in a heat exchanger (fig. \ref{fig:fan_heat_exchanger}). In most cases, the air is sucked by the fan through the heat exchanger which creates a turbulent inflow that interacts with the fan blades and generates low frequency sound. In order to localize sound sources at low frequencies, large microphone arrays have to be used. The sound propagation from the fan to the microphones is then influenced by the reflection and shading by the housing of the heat exchanger.

If in such cases beamforming is done based on free field Green's functions, the disturbance of real sound path leads to insufficient beamfoming results in the lower frequency range, where the outer (in this case shadowed) microphones are required for a sufficient source localisation. The aim of the present work is to develop a method to calculate the Green’s functions for typical fan installation conditions and to use these tailored Green's function in the calculation of the beamforming steering vectors. 

The aim of this paper is not to consider movement of the sound sources yet but to determine and apply tailored Green's functions for stationary beamforming. Furthermore the flow is not considered. Sarradj et al. \cite{sarradj2020efficient} have shown that the influence of the intake flow on the sound propagation can be neglected for a slow moving fan (tip speed $<0.15\, Ma$).

In the following, an overview of beamforming methods under disturbed sound propagation will be given. So far these have been mainly investigated for closed test sections of wind tunnels.

The first published method is the Image Source Method (ISM) by Guidati et al. \cite{guidati2002reflection}. The positions of reflected mirror sources are determined numerically and taken into account in the steering vectors. The damping of each mirror source was determined experimentally by measurements in the reverberant environment of a closed test section.  Using the same idea, Fischer \& Doolan \cite{fischer2017beamforming} presented two methods, numerically and experimentally, to involve the reflections in beamforming in closed test sections. The numerical method is also based on Green's functions with mirror sources according to the wind tunnel wall reflections. The second method measured the Green's functions with an generic source inside the closed test section at the focus grid points. The source localization was mostly improve when using the experimental Green's function, especially at high frequencies. The best results were obtained when using the experimentally determined Green's function combined with CLEAN-SC. 

In \cite{sijtsma2003corrections}, Sijtsma and Holthusen have presented an approach in which the beamformer output is minimized for the directions of the mirror sources. They added a control mechanism to preserve robustness especially at low frequencies. This method succeeded well in reconstructing source levels in the entire frequency range investigated in the paper. Below $f = 2500\,$Hz however, the position of the maximum in the beamforming map deviates significantly from the actual source position.

Another approach was published by Fischer \& Doolan \cite{fischer2017improving}. They identified the mirror sources by means of different time delays in the cross-correlation matrix (CCM) and removed their influence by multiplying each CCM vector by an window function centered at the main peak. The cross-correlation matrix is derived from the cross-spectral matrix. An advantage of the method is that no information about the geometry of the reverberant environment is required. It could be shown that the beamforming maps in a reverberant test section could be improved for a loudspeaker without flow as well as for an airfoil in a flow. The beamforming results for the loudspeaker were comparable with the theoretical Point Spread Function of the array.

Bousabaa \cite{bousabaa2018acoustic} presented a method to determine the Green's functions numerically in the time domain. Using a single aeroacoustic CAA simulation, the Green's functions are calculated from all focus points to all microphones. Therefore, both flow and diffracting objects are taken into account in wind tunnel measurements. The method takes advantage of the sparsity of Green's functions for these cases. Therefore, only external aeroacoustic problems can be modelled, because in other cases the Green's functions are not sparse enough. The method is therefore only suitable for non-reflective measurement environments.

A further simulation-based approach is presented by Kaltenbacher et al. \cite{kaltenbacher2018inverse}. Here, first the acoustic wave equation is solved using the Finite Element Method (FEM). In doing so, the actual boundary conditions of the given measurement setup can be considered. Second, the inverse problem is solved to match both, the measured and simulated pressure. The goal is to consider acoustic problems in the low frequency range. In numerical 2D and 3D setups it can be shown that the method improves sound source localization at low frequencies \cite{kaltenbacher2018inverse, gombots2018inverse}. Additionally, in \cite{gombots2016combined}, the Green's functions from the FEM have been used directly to calculate the steering vectors for numerical beamforming. This use of numerically calculated steering vectors improved the results compared to beamforming maps with free-field steering vectors as well. The calculation of Green's functions is done reciprocal, i.e. from the microphones to the focus points. Thus, the number of simulations to be run corresponds to the number of microphones.

Most of the presented methods have been developed for closed test sections in wind tunnels and therefore only consider reflections. The method of Bousabaa however, has been developed for non-reflective environments. For the presented fan application with heat exchanger, the Green's functions includes reflection, diffraction and shadowing effects. This is necessary to use a microphone array whose aperture is larger than the cross section of the heat exchanger. The method of Kaltenbacher determines the complete Green's functions by FEM simulations. However, its applicability has so far been shown only for numerical cases. 

In the present work, simulation based Green's functions shall be used for beamforming based on real measurement data. The simulations are done using the Boundary Element Method (BEM). The Green's functions consider the effects of reflection, diffraction and shadowing for arbitrary geometries. Beamforming is applied in the frequency domain using different steering vector formulations. The steering vector formulations are known to find for free-field Green's function either the actual source position as a local maximum or the true source power at the source position, see Sarradj \cite{sarradj2012three}). As this was unknown for the use  of  general Green's functions, a parameterized and generalized steering vector formulation was developed. For this generalized formulation it was proven that by means of choice of parameter, the beamforming maps will show the desired properties (see section \ref{section:appendix_1} "appendix").

The paper is organized as follows:

First, in section \ref{sec:methods} the beamforming formulations are presented. The different steering vector types are also considered. In the appendix a generalization of the common steering vector formulations and simplified characterizations of their properties are given. Then, criteria for the evaluation of the beamforming results are introduced and the investigated test case is presented. In section \ref{sec:FD} the beamforming results are presented and discussed. The work is summarized in section \ref{sec:conclusion}.\\

\section{Methods}
\label{sec:methods}
In the following chapter the basics for the application of the beamfoming with BEM simulated Green's functions will be given. First of all, an overview of the necessary mathematical formulations of beamforming is given including the calculation of different beamfoming steering vectors using numerically calculated Green's functions. Afterwards, different validation criteria for the evaluation of the beamforming results are introduced. Lastly, the experimental test case selected for validation is presented.

\subsection{Beamforming} Beamforming of phased microphone array measurements is a technique for sound source localization and the estimation of the according source power. Therefore, given focus points are scanned at locations where possible sound sources are expected. This is done by manipulating the microphone signals for each focus point in order to compensate for time delay and attenuation of amplitude between the focus point and each microphone position. Afterwards, all microphone signals are summed up and the result is recorded in a source map. This source map is often called \textit{dirty map}, because it includes all the influences of the used microphone array and steering vectors. Often, this \textit{dirty map} is \textit{cleaned} using further methods called \textit{deconvolution} methods. These deconvolution methods are however not in the focus of this paper.

The beamforming procedure can be performed either in the time domain or in the frequency domain. Both approaches will be described in the following based on \cite{sijtsma2004experimental}.
\subsubsection{Time domain beamforming} In the time domain, the estimated source signal $\bar{\sigma}\left(t\right)$ for a focus point $\mathbf{y}$ can be calculated by:
\begin{equation}
\bar{\sigma}(t,\mathbf{y}) = \frac{1}{M} \sum_{m=1}^{M} 4\pi \vert \vert	 \mathbf{x}_m - \mathbf{y} \vert \vert p_m \left( t+ \Delta t_{m} \right).
\label{eq:TD_BF}
\end{equation}
In this formulation, the time signal of the m-th microphone $p_m$ is shifted by the time delay $\Delta t_{m}$ estimated for the distance between the microphone position $\mathbf{x}_m$ and the focus point $\mathbf{y}$:
\begin{equation}
\Delta t_{m} = \frac{\vert \vert \mathbf{x}_m-\mathbf{y} \vert \vert}{c},
\end{equation}
where $c$ is the speed of sound.

For deeper evaluation of the results, $\bar{\sigma}(t,\mathbf{y})$ can be transformed into the frequency domain. For this, the power spectra $\hat{\sigma}(f,\mathbf{y})$ are calculated by performing a Fourier transform to the signals at each focus point signals using Welch's method \cite{welch1967use}.

\subsubsection{Frequency domain beamforming using cross spectra} Another approach is to apply beamforming to the microphone signals in the frequency domain. A computationally fast approach to this is to separate the processing of the microphone data from the calculation of the sound propagation information. This way, only one transformation of the microphone signals to the frequency domain has to be performed. The propagation from a every possible chosen source position to each microphone is performed separately.
The complex pressure amplitudes $\hat{p}_m(f)$ of the microphone signals can be calculated by a discrete Fourier transformation for a block of K samples:
\begin{equation}
\hat{p}_m(f) = \frac{2}{K} \sum_{k=1}^{K} p_{m,k} e^{-2 \pi ifk \Delta t}.
\end{equation}
To estimate the degree of correlation between two microphone signals, the cross power is calculated:
\begin{equation}
C_{mn}(f) = \frac{1}{2}\hat{p}_m(f)\hat{p}_n^*(f).
\end{equation} 
where $*$ indicates the conjugate complex.
This is done using Welch's method \cite{welch1967use}. In the following the dependency on the frequency $f$ is mostly omitted to increase the readability. The auto- and cross-powers of all microphone signals are stored in the cross-spectral matrix (CSM)  $\mathbf{C}$:
\begin{equation}
\mathbf{C} = \frac{1}{2} \hat{ \mathbf{p}} \hat{ \mathbf{p}}^* \in \mathbb{C}^{M\times M},
\end{equation}
with the $M$-dimensional vector $\hat{ \mathbf{p}}$ of complex pressure amplitudes:
\begin{equation}
\hat{ \mathbf{p}}=\left( \begin{array}{c}
\hat{ p}_1(f) \\
. \\
. \\
. \\
\hat{p}_M(f) \\
\end{array} \right)
\end{equation}
For the estimation of the source power at a certain focus point, information about the transfer function or Green's function $g(\mathbf{x},\mathbf{y},\omega)$
\begin{equation}
        g: \mathbb{R}^3 \times \mathbb{R}^3 \times \mathbb{R} \rightarrow \mathbb{C}
\end{equation}
from the focus point position $\mathbf{y} \in \mathbb{R}^3$ to the microphone position $\mathbf{x} \in \mathbb{R}^3$ is needed. This information is stored in the vector of the Green's functions $\mathbf{g} = \mathbf{g}(\mathbf{y})\in \mathbb{C}^M$, where $g_m=g_m(\mathbf{y})=g(\mathbf{x}_m,\mathbf{y})$ and $\mathbf{x}_m \in \mathbb{R}^3$ is the position of the $m$-th microphone. Its components are the pressure amplitudes induced by a unit sound source at the focus point in dependence of the source model and the radiation conditions. For a monopole source and free-field conditions, $g_m$ for a certain microphone $m$ can be written as follows:
\begin{equation}
g_m = \frac{1}{r_m} e^{-jkr_m},\quad r_m = \|\mathbf{y}-\mathbf{x}_m\|_2,
\label{eq:h_FF}
\end{equation}
where $r_m$ is the distance between the focus point $\mathbf{y}$ and the $m$-th microphone position $\mathbf{x}_m$. In this paper, the Green's functions $g_m$ for the \textit{axial fan with heat exchanger - test case} are calculated numerically by boundary element simulations. 

To estimate the source auto-power at a certain focus point one has to evaluate:
\begin{equation}
A = \mathbf{w^* C w}
\label{eq:beamfoming}
\end{equation}
where $\mathbf{w}$ is the so called steering vector, which is defined in the next section.
\subsubsection{Steering vector formulations}
\label{subsubsec:Sarradj}
There are several possible formulations for the steering vector $\mathbf{w} = \mathbf{w}(\mathbf{y}) \in \mathbb{C}^M$. In this work, three of them are applied on analytical free-field and on numerically computed arbitrary Green's functions. The formulations and the numbering of the steering vector formulations is done analogously to Sarradj in \cite{sarradj2012three}. 

\textbf{Formulation I:} For the first formulation, only the phase part of the Green's functions $\mathbf{g}$ is considered:
\begin{equation}
w_m^{I} = \frac{1}{M} \frac{g_m}{\vert g_m \vert}
\label{eq:SarradjI}
\end{equation}
where $M$ is the number of microphones. This formulation meets the condition that a local maximum output power of the beamformer is provided at the source location.

\textbf{Formulation II:} The second formulation also compensates for the amplitude reduction in free-field radiation. It is the frequency domain formulation of eq. \ref{eq:TD_BF}:
\begin{equation}
w_m^{II} = \frac{1}{M} \frac{g_m}{\vert g_m \vert^2}
\label{eq:SarradjII}
\end{equation}
The source power is reconstructed correctly, but the source position may not correspond to a local maximum of the beamforming map.

\textbf{Formulation III:} Another way to formulate the steering vector is to minimize the difference between the measured pressure vector $\mathbf{p}$ and the product of the complex source amplitudes $a$ and the Green's functions $\mathbf{g}$:
\begin{equation}
J = \vert \vert \mathbf{p}-a\mathbf{g} \vert \vert^2
\end{equation}
The resulting steering vector formulation also reconstructs the source power correctly, but again not the source position at a local maximum of the beamforming map:
\begin{equation}
w_m^{III} = \frac{g_m}{\|\mathbf{g}\|_2^2}
\label{eq:SarradjIII}
\end{equation}
In the special case of a free field without flow Green's function (eq. \ref{eq:h_FF}) the steering vector formulations are:
\begin{equation}
    \begin{split}
        w_m^{I} &= e^{-jkr_m} \frac{1}{M}\\
        w_m^{II} &= e^{-jkr_m} \frac{r_m}{M}\\
        w_m^{III} &= e^{-jkr_m} \frac{1}{r_m} \frac{1}{\sum_m r^{-2}_m}
    \end{split}
\end{equation}

In appendix \ref{section:appendix_1} a unified steering vector formulation is presented including the above three types $w_m^{I}$,$w_m^{II}$ and $w_m^{III}$ as special cases. A proof is given that also for arbitrary differentiable Green's functions $g_m$ the requirement for the correct amplitude is fulfilled only for steering vector formulation II and III and the requirement for a local maximum of the beamforming response is fulfilled only for steering vector formulation I.

\subsubsection{Point spread function} The point spread function is the spatial impulse response of an array on a unit point source at a focus point $\mathbf{y}_s$. As it induces a CSM by
\begin{equation}
\mathbf{C}_s = \mathbf{g}(\mathbf{y}_s) \mathbf{g}(\mathbf{y}_s)^*
\end{equation}
the source power can be calculated as follows:
\begin{equation}
A^{\text{PSF}}(\mathbf{y}) = \mathbf{w}(\mathbf{y})^* \mathbf{C}_s \mathbf{w}(\mathbf{y}) = \mathbf{w}(\mathbf{y})^* \mathbf{g}(\mathbf{y}_s) \mathbf{g}(\mathbf{y}_s)^* \mathbf{w}(\mathbf{y}) = \left|\mathbf{w}(\mathbf{y})^* \mathbf{g}(\mathbf{y}_s)\right|
\label{eq:PSF}
\end{equation}

\subsection{Green's functions from boundary element method simulations for steering vector calculation}
In this paper the Green's functions are calculated with the boundary element method (BEM) in the frequency domain. The simulation setup will be described in section \ref{subsubsec:sim_set}. The calculation is done reciprocal which means that the scattering problem of a Delta function $\delta$ located at the microphone position $\mathbf{x}_m$ is solved for each microphone. 
\begin{equation}
g^{\text{BEM}}_m(\mathbf{y})= g(\mathbf{y}, \mathbf{x}_m)
\label{eq:h_BEM}
\end{equation}
Due to the known reciprocal property of the wave equation without flow and arbitrary boundary conditions \cite{rayleigh1896theory}, this is the same as calculating the Green's function from the focus point $\mathbf{y}$ to the microphone position $\mathbf{x}_m$
\begin{equation}
    g(\mathbf{y}, \mathbf{x}_m) = g(\mathbf{x}_m, \mathbf{y})
\end{equation}
The solution is evaluated at the focus points $\mathbf{y}$ of the beamforming map. The resulting complex pressure amplitudes can be used directly as Green's functions $g_m$ in the different steering vector formulations (eq. \ref{eq:SarradjI}, \ref{eq:SarradjII} and \ref{eq:SarradjIII}) for frequency domain beamforming.

\subsection{Validation criteria}
\label{subsec:val_crit}
The beamforming results computed with the different steering vectors and four generic source positions will be compared in the next section. Therefore, five different validation criteria are now established as measure of the resulting beamforming maps. It shall be mentioned again that all criteria are based on so-called 'dirty maps' on identical grids without further deconvolution methods. These criteria shall express different properties of the beamforming maps in single values. They aim at two aspects: The first is whether the sound source is found correctly with regard to both, position and amplitude. The second one is how other sources are masked by a broad main lobe or side lobes of the point spread function.

The definitions of the first and second criterion are based on Sarradj \cite{sarradj2012three} and aim at the correct reconstruction of the sound source: First, the location of the maximum should match the actual source position $\mathbf{y}_s$. The criterion is therefore the \textbf{spatial deviation} of the location of the maximum from the source position:
\begin{equation}
\vert \vert \Delta \mathbf{y} \vert \vert = \vert \vert \mathbf{y}_{\text{max}} - \mathbf{y}_s \vert \vert
\label{eq:conditionI}
\end{equation}
Second, the beamformer output at the source position should be a measure of the \textbf{source strength}, as the difference between the indicated source strength $ A_e$ and the true source strength $A_s$ 
\begin{equation}
\Delta L = A_e -A_s. 
\label{eq:conditionII}
\end{equation}
The true source level corresponds to the point spread functions (eq. \ref{eq:PSF}), which means  that the true source strength must be equal to one:
$ A(\mathbf{y} = \mathbf{y}_s) = 1 $

According to reference \cite{sarradj2012three} and the appendix in section \ref{section:appendix_1}, none of the steering vector formulations from section \ref{subsubsec:Sarradj} fulfills both criteria exactly. Nevertheless, in the appendix it is shown that both criteria can be fulfilled up to an arbitrary small error. Unfortunately this is not a practical approach for application cases, as sidelobes are clearly increased as well. Formulation I fulfills the first criterion and formulation II and III fulfill the second criterion.

The second major aspect of the validation -the masking of other sources by the point spread function - will be based on the third and fourth criterion. The third criterion is the \textbf{spatial resolution} $b$. It is a measure of how large the distance between two sound sources must be in order to be separated by the beamformer. In the present work it is calculated as twice the distance between the position of the maximum $\mathbf{y}_{\text{max}}$ and the point on the $L_{\text{max}}-1\,$dB contour line most distant from the maximum. The use of the $L_{\text{max}}-1\,$dB line instead of the usually used $L_{\text{max}}-3\,$dB line is justified by the fact that at low frequencies the resolution could otherwise not be determined due to the large main lobe widths.
\begin{equation}
b = 2 \cdot \max_{\hat{\mathbf{y}}\in S} \left( \vert \vert \mathbf{y}_{\text{max}} - \hat{\mathbf{y}} \vert \vert \right), \quad S =\{\mathbf{y}|L_{\text{max}}-1 = \text{dB}(A(\mathbf{y}))\}
\end{equation} 
The fourth criterion is the \textbf{main lobe to side lobe ratio (MSR)}. It is also often referred to as dynamics. In this paper it is calculated as the difference between the level at the source position $L_s$ and the level of the highest side lobe $L_{SL}$:
\begin{equation}
MSR = L_s - L_{SL}
\end{equation}
The highest side lobe is determined iteratively: Starting from the maximum of the beamforming map, the level is incrementally reduced. At each iteration the corresponding isolines are drawn into the map. At first, only the area of the main lobe is found. As soon as another area outside the main lobe is found, the highest side lobe is assumed there.

The fifth and final criterion, the \textbf{source to pattern ratio (SPR)}, addresses both aspects. It takes into account that not only the level of the highest side lobe, but the total energy added to the map by the point spread function is important for the quality of the results. Therefore, it is defined as the level of the ratio between the beamformer output at the source position and the average output at the other focus points:
\begin{equation}
SPR = 10 \log_{10} \left( \frac{ A \left(\mathbf{y}_s \right)}{\frac{1}{ N_{map}}\sum_{n=1}^{N_{map}} A \left( \mathbf{y}_n \right)} \right)
\label{eq:SPR}
\end{equation}
where $\mathbf{y}_n$ is the $n$-th focus point.

\subsection{Validation setup}
\label{subsec:val_setup}
The following section will describe the setup that is used to validate the method. First the measurement setup and then its implementation in the BEM simulation will be presented.     
\subsubsection{Measurement Setup}
\label{subsec:meas_setup}
The  goal of the presented method is to improve \textit{dirty map}- beamforming results for sound source localisation of axial fans mounted in heat exchangers. A simplified setup is shown in figure \ref{fig:meas_setup}. A loudspeaker is used as generic sound source because the presentation of beamforming using tailored Green's functions is in this paper restricted to time invariant sound sources and transfer functions. As a further simplification, the heat exchanger is modeled as a cuboid box without condenser.

The loudspeaker is a NTi Talkbox \cite{2020talkbox} emitting white noise between $f = 100\,$Hz and $f = 8\,$kHz as test signal. Measurements are done for four different positions of the loudspeaker and the positions are shown in figure \ref{fig:sPos}.

The ebm-papst testbench for combined air and sound measurements is used for the validation experiments. It consists of two half-anechoic chambers separated by a wall where usually the fan is mounted in a so-called wall ring. The cuboid box is arranged concentric with the wall ring. It is connected to the wall and has the dimensions $1.54 \times 1.14 \times 0.46 \, $m$^3$. 

The microphone array is arranged in two circles with $d_1 = 1.6\,$m and $d_2 = 0.8\,$m diameter. The larger circle houses $M = 40$ and  the smaller circle $M = 24$ microphones. It is also arranged concentric with the wall ring. The distance from the separating wall to the larger microphone array circle is $z = 0.8\,$m and $z = 1.3\,$m to the smaller circle.

The data acquisition system is an I$^2$S- frontend by CAE Software and Systems GmbH with the corresponding digital MEMS microphones. The signals were recorded for 15 seconds, with sampling rate $f_s = 48077\,$kHz.

As the standard frequency beamforming procedure, in the first step the cross-spectral matrix $C$ of the data was calculated using Welch's method for estimating the cross-spectra. A Hann weighting window was applied to each subset of the data.

\begin{equation}
	C_{nm} = \sum_{k=1}^{N}{\left(w\cdot\phi_n^k\right) \cdot \left(w\cdot\phi_m^k\right)^H}
\label{eq:Rmatrix}
\end{equation}

\noindent where $w$ indicates the Hann weighting function, $H$ indicates the Hermitian, and $\phi_n^k$ is the Fourier transform of the $k$-th window of the signal of microphone $n$ and of microphone $m$, respectively.

To ensure that the frequency lines in the CSM match the frequencies of the simulated, tailored Green's functions, a window size of $L = 4809$ samples was used to achieve a frequency spacing of $\Delta f = 10\,$Hz. The total number of $N = 298$ averages was achieved using an overlap factor of $r = 0.5$. The procedure results in a matrix $C$ for every bin center frequency of the Welch estimation.

\subsubsection{Simulation Setup}
\label{subsubsec:sim_set}
The Green's functions are computed numerically  with the Fast Multipole-Boundary Element Method (FM-BEM) code FMCAS \cite{lummer2019installation,lummer2013validation}. FMCAS solves the Helmholtz equation using a boundary integral equation discretized on a triangulated surface. The function to be determined is assumed to be constant on each triangle. The resulting system of linear equations is solved using an iterative solver from the PETSc library. The computation of the matrix-vector products is accelerated by a high-frequency Fast Multipole Method (FMM) based on a plane wave approximation.
Further details of FMCAS can be found in the references cited.

In the simulations only the reverberant areas of the test bench are considered. These are the floor and the central part of the separating wall which are modelled fully reflective. 

The surfaces covered with absorbers are modeled as free field radiation. This applies to the ceiling and most parts of the walls.

\begin{figure}
	\centering
	\begin{subfigure}[b]{0.3\textwidth}
		\centering
		\includegraphics[width=\textwidth]{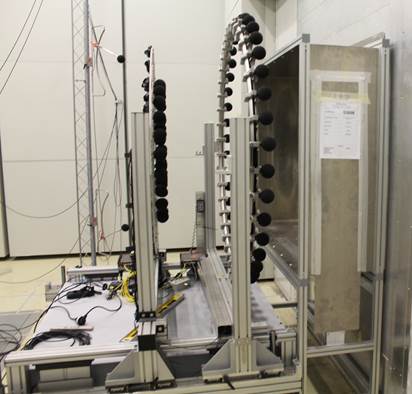}
		\caption{}
		\label{fig:meas_setup}
	\end{subfigure}
	\begin{subfigure}[b]{0.3\textwidth}
		\centering
		\includegraphics[width=\textwidth]{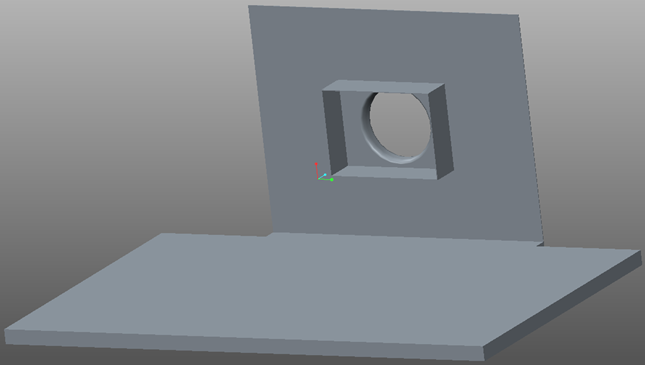}
		\caption{}
		\label{fig:sim_setup}
	\end{subfigure}
	\begin{subfigure}[b]{0.3\textwidth}
		\centering
		\includegraphics[width=\textwidth]{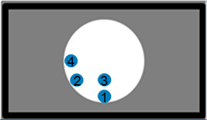}
		\caption{}
		\label{fig:sPos}
	\end{subfigure}	
	\caption{Validation setup. (a) Side view of the measurement setup. (b) Simulation setup. (c) Source positions.}
	\label{fig:setup}
\end{figure}

Furthermore, only one of the two rooms of the test bench is considered. From that, only half of the floor area is modeled. A CAD model is shown in figure \ref{fig:sim_setup}.

The cell size of the simulation grid is determined as a function of the simulated frequency. Above $f = 1000\,$Hz grids with 6 points per wavelength (ppw) are used. In a preliminary study \cite{ruck2019entwicklung} it was shown that below $f = 1000\,$Hz this is not sufficient to resolve the geometry properly. Therefore, a frequency-independent grid is used in this range, which has a resolution of 12 ppw at $f = 1000\,$Hz. At the lowest simulated frequency $f = 120\,$Hz the resolution is thus 100 ppw.

Simulations are carried out for frequencies between $f = 120\,$Hz and $f = 2040\,$Hz. The distance between two frequencies is maximum $\Delta f = 120\,$Hz. Between $f = 480\,$Hz and $f = 1080\,$Hz simulations for additional frequencies are made to be able to investigate this frequency range in more detail.

The Green's functions are evaluated on a square grid at a distance of $z = -0.03\,$m in front of the separating wall to avoid numerical effects close to the wall. It has the dimensions $1.44 \times 1.44\,$m$^2$. The distance between two evaluation points is $dxy = 0.01\,$m. This grid is also used as focus grid for the beamforming evaluations.

\section{Results}
\label{sec:FD}
In the following chapter, the method of tailored Green's functions will be used for beamforming on the presented test case. The beamforming results calculated all in the frequency domain with different steering vectors will be tested using the validation criteria introduced above. 

The algorithm is first validated for a synthetic test case of the setup introduced in section \ref{subsec:val_setup} where both, Green's functions and microphone data are based on simulation. Afterwards the different steering vectors based on the simulated Green's function, are applied to data from a real measurement. The results will be discussed at the end of this section.

For the evaluation of the validation criteria only focus points inside the box are considered.

\subsection{Point spread functions}
\label{subsec:PSF}
In the first step the beamforming algorithm based on BEM simulated Green's function is validated and the performance compared with standard beamforming.  Therefore, the point spread functions (PSF) of the microphone array will be evaluated using the three different steering vectors based on free field and BEM simulated Greens's functions, respectively. The PSF, which is the source map of a signal unit source, is calculated according to eq. \ref{eq:PSF}. To simulate a source at a certain loudspeaker position, the CSMs are calculated from BEM computed Green's functions for the corresponding source position - microphone array combination. The aim of this first step is to verify whether the first and second validation criteria (see section \ref{subsec:val_crit}) are fulfilled when the correct (tailored) Green's functions are used. Therefore, the steering vectors are calculated according to formulation I (eq. \ref{eq:SarradjI}) and formulation III (equation \ref{eq:SarradjIII}) from both free-field and BEM computed Green's functions. The results  for the steering vector formulation II (eq. \ref{eq:SarradjII} will follow in section \ref{subsec:results_form_II}.

The value of the beamforming validation criteria are derived from the individual beamforming maps of four different source positions with the same grid according to section \ref{subsec:val_crit}. Then, the average of the different criteria for the four source positions are calculated respectively. Sample beamforming maps are shown in the appendix (fig. \ref{fig:Maps_PSF}).

\subsubsection{Spatial deviations} 
\begin{figure}
	\centering
	\begin{subfigure}[b]{0.45\textwidth}
		\centering
		\includegraphics[width=\textwidth]{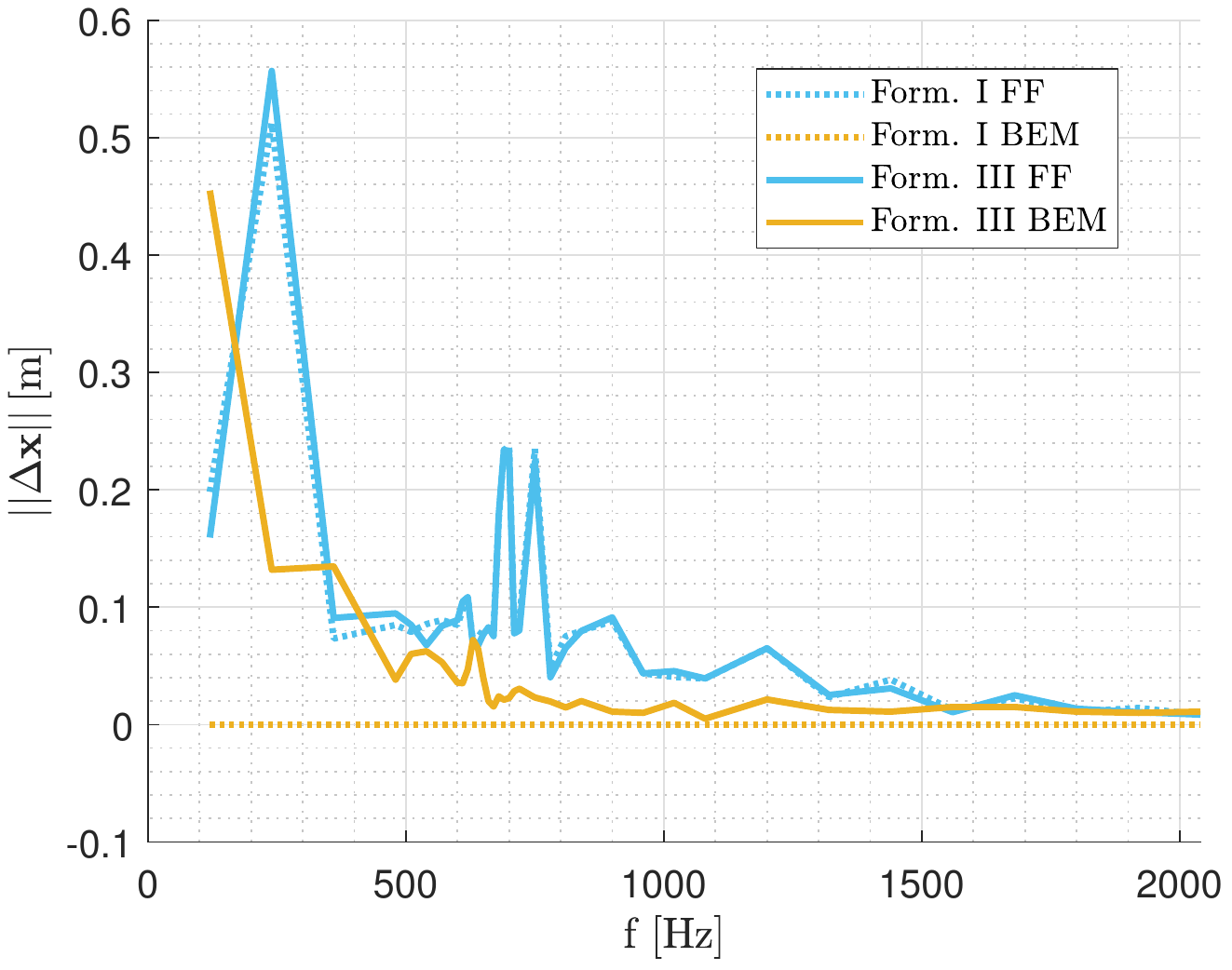}
		\caption{}
		\label{fig:dx_MeanPSF}
	\end{subfigure}
	\begin{subfigure}[b]{0.45\textwidth}
		\centering
		\includegraphics[width=\textwidth]{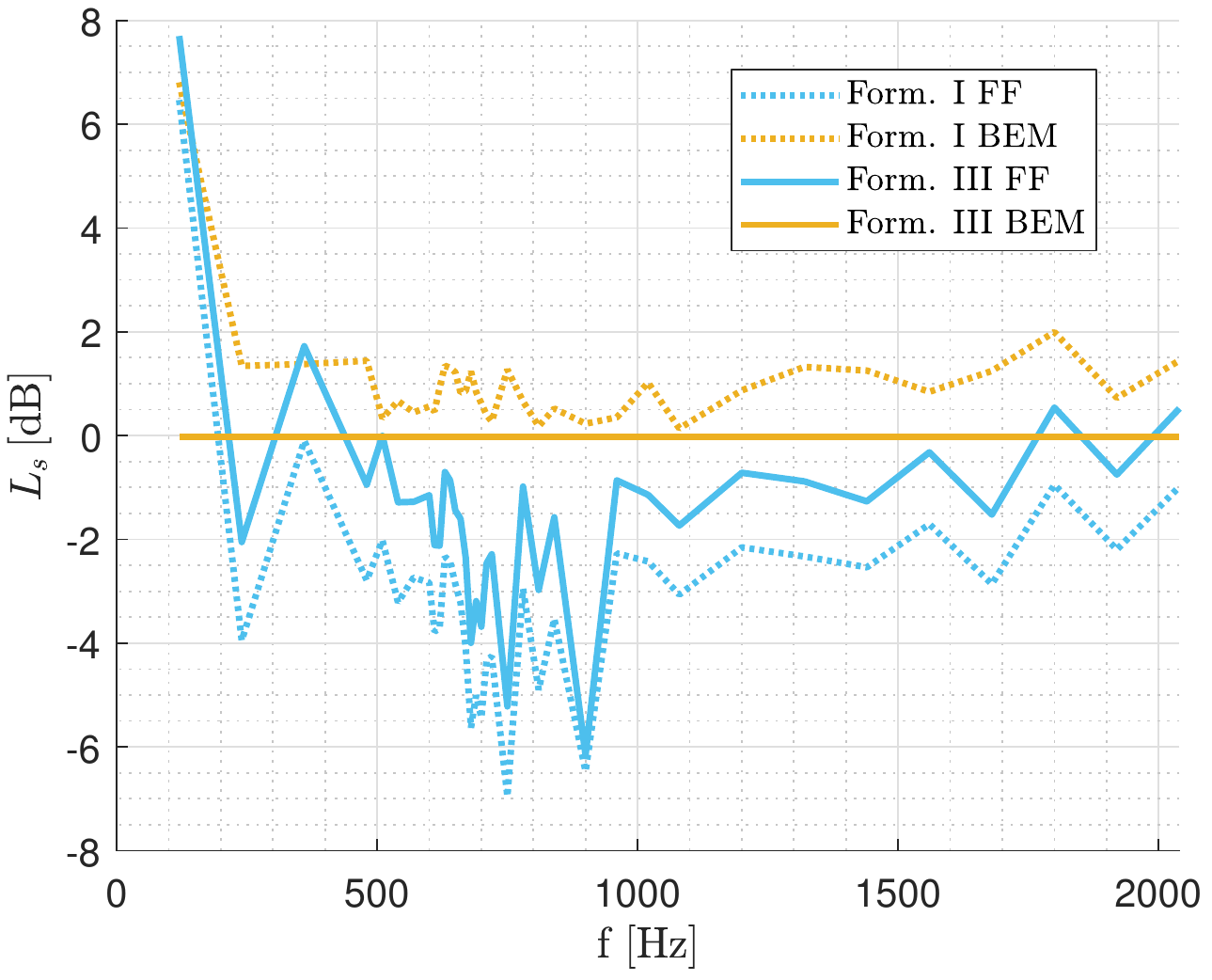}
		\caption{}
		\label{fig:Ls_MeanPSF}
	\end{subfigure}	
	\caption{Results for PSFs from BEM generated Green's functions. Spatial deviation of the maximum from the source position $\vert \vert \Delta \mathbf{y} \vert \vert$ (a) and estimated difference of the sound pressure level $\Delta L_s$ $\Delta z = 1\,$m away from source (b) over frequency for free-field (FF) and BEM-calculated (BEM) Green's functions and vector formulations I and III. Average values over four source positions.}
	\label{fig:PSF_crit}
\end{figure}

Figure \ref{fig:PSF_crit} shows on the left side the average spatial deviation between the maximum of the beamforming map and the actual source position in the frequency range from $f = 120\,$Hz to $f = 2040\,$Hz, for the steering vectors I and III based on the two different Green's functions, respectively. 

For the steering vectors calculated from the BEM-simulated Green's functions according to formulation I , the spatial deviation of the maximum from the actual source position is $\vert \vert \Delta \mathbf{y} \vert \vert = 0\,$m for all frequencies (see fig. \ref{fig:dx_MeanPSF}).  The position of the maximum corresponds in this case to the actual source position and the first criterion is fulfilled. The source positions calculated with the same steering vector formulation I but based on the free-field Green's function deviate considerably from the true source position with the deviation increasing with wavelength.

When using the BEM-steering vectors calculated according to formulation III, the averaged deviation is $\vert \vert \Delta \mathbf{y} \vert \vert \neq 0\,$m at all frequencies. Nevertheless, the deviations are still considerable smaller for almost all frequencies than for the cases calculated with free-field Green's functions. 

The difference between the spatial deviation of the source position based on the two calculation with the free-field Green's function are only very small. Overall, for all three formulations with deviating source positions, the deviation nearly vanishes at higher frequencies, $ f> 1500\,$Hz. 

\subsubsection{Estimated source strength}
Figure \ref{fig:PSF_crit} shows on the right side the sound pressure levels at a distance of $z = 1\,$m from the source in the frequency range from $f = 120\,$Hz to $f = 2040\,$Hz.

The true source power is only to be found by the beamforming calculations using steering vector formulations III based on the BEM- simulated Green's function. Again, this finding is in accordance with the theoretical analysis in the appendix. 

The same steering vector formulation III, but based on the free field Green's function leads to various deviating source strength up to $\pm6\,$ dB in the considered frequency range.

The source power calculated with the steering vector formulation I, based on BEM simulated Green's functions, deviates above $300\,$Hz approx. $+1\,$ dB from the true source power. Also, the source power calculation with the steering vector formulation I, based on free field Green's functions deviates approx. $1\,$dB more than the according steering vector formulation III.

\subsubsection{Conclusions from PSF results}
The evaluation of the point spread functions has shown that for the presented test case the use of the \textit{correct} tailored Green's functions for steering vector calculation improves the results in the considered frequency range. When using the free-field steering vectors for the beamforming calculation neither the correct source position nor the correct source level could be determined. The deviations increase especially for frequencies below $f = 1000\,$Hz.
Apparently, the simplification using the free field propagation model for the steering vector calculation does not match the actual sound propagation.

If, on the other hand, the matching tailored Green's functions are used, the corresponding criterion depending on the steering vector is fulfilled. However, in the presented case the steering vector formulation has a greater influence on the deviation from the actual source position than with free-field Green's functions. 

From the evaluation of the point spread functions it can be concluded that the presented method is validated and significantly improves the beamforming results for the test shown. Source position and strength are found properly, when using the correct Green's function and the corresponding steering vector formulation.

\subsection{Measurement results for steering vector formulation I and III}
Now that the method has been validated for a synthetic test case in the previous section, it will be investigated how well the simulated Green's functions are suited for application to real microphone array measurements.

Microphone array measurements were conducted with a loudspeaker as sound source as described in section \ref{subsec:meas_setup} at the same four positions as those used for the synthetic case the previous section. 

The steering vectors are calculated as in section \ref{subsec:PSF} on the basis of free-field or BEM Green's functions, respectively. The beamforming maps are subsequently calculated using equation \ref{eq:beamfoming}.

In figure \ref{fig:FD}, the quality of the beamforming maps calculated with steering vector formulation I and III and the free field and BEM Green's function are evaluated using four validation criteria from section \ref{subsec:val_crit}. The criterion \textit{estimated source strength} is not considered here because the actual source strength of the loudspeaker is not known exactly. Again, the average of the results at the four source positions is shown as a function of frequency. In addition, the curves for the synthetically generated PSFs with steering vector formulation I and III based on  BEM-simulated Green's function given as a reference. 

Furthermore, in figure \ref{fig:Maps_FD}, some sample beamforming maps are shown exemplary to illustrate and explain the differences of the beamforming results. The beamforming maps are calculated using formulation III of free-field and BEM steering vectors. The true source positions are marked with a cross. It should be noted that in figure \ref{fig:FD} the average values of the validation criteria (of all four measurements with the four different source positions) are shown, but the beamforming maps in \ref{fig:Maps_FD} are only the result of one source position.

The results in this section will first be presented and afterwards discussed in section \ref{subsubsec:Meas_res}.

\subsubsection{Spatial deviations}
Figure \ref{fig:dx_MeanFD} compares the spatial deviations $\vert \vert \Delta \mathbf{y} \vert \vert$ between the maximum of the beamforming map and the true source position. The accuracy of the measured source position increases with the frequency for all steering vector formulations. The deviation is not $\vert \vert \Delta \mathbf{y} \vert \vert = 0\,$m in any case, but it is the smallest for the BEM steering vector generated according to formulation I. In this case, the deviation between beamforming map maximum and true source position is smaller than $5\,$cm in the frequency range above $f=250\,$Hz. Even using formulation III, the results for most frequencies improve by using the BEM- simulated Green's function instead of the free-field Green's functions. Especially outliers that occur in the lower frequency range at $f < 1\,$kHz exhibit a smaller spatial deviation when using the BEM- simulated Green's function. Above this frequency, the deviation for steering vector formulation I and III of BEM- simulated Green's functions is the same for almost all frequencies.

The beamforming maps of formulation III, shown for the free-field Green's functions and the BEM-simulated Green's functions in figure \ref{fig:S1_660Hz_FD_FF_steering_Conventional} and \ref{fig:S1_660Hz_FD_BEM_steering_Conventional} respectively for $f = 660\,$Hz, support these observations. There is no maximum at the source position in the free-field evaluation, but several lobes are distributed over the map. When using BEM steering vectors, the main lobe is located just above the actual source position and has clearly the highest level.

It appears that the reflection caused by the solid walls of the heat exchanger are interpreted as additional source outside the heat exchanger. The proportion of undisturbed sound propagation from the source to the array microphones is too small for a localization of the true source position.   

\subsubsection{Resolution}
In figure \ref{fig:resolutionHhMax_MeanFD} the spatial resolution $b$ of the beamforming measurements with the different steering vectors is shown and again compared to the spatial resolution of the PSF simulations. The spatial resolution $b$ is normalized with the wavelength $\lambda$ in order to investigate whether there is a Helmholtz similarity. As mentioned in section \ref{subsec:val_crit}, the main lobe width in this paper is defined as the $L_{\text{max}}-1\,$dB line instead of the usually used $L_{\text{max}}-3\,$dB line.

The resolution of the beamforming maps based on the free-field steering vectors is approximately proportional to the wavelength for most frequencies. The ratio $b/\lambda$ usually takes values between $0.4<b/\lambda<0.45$. Below $f=800\,$Hz some outliers are found, especially for formulation I. Results based on Green's functions from BEM simulations show less proportionality to wavelength, which means especially for frequencies below $f=1200\,$Hz that the main lobe width is reduced by using the BEM-simulated Green's functions. At low frequencies $f<500\,$Hz, the resolution is up to $\Delta b/\lambda = 0.25$ which is considerably higher resolution than with the corresponding free-field result. Exemplary beamforming maps for $f = 570\,$Hz and $f = 660\,$Hz are shown in figure \ref{fig:Maps_FD}, illustrating the significantly broader mainlobes for the free-field case.

Above $f = 1200\,$Hz, all main lobe widths of free-field and BEM cases are usually comparable within a deviation of $\Delta b/\lambda = \pm 0.06$, with the exception at $f = 1560\,$Hz, where the beamforming resolution based on the BEM-simulation Greens's functions deviates by $\Delta b/\lambda = 0.3$.

\subsubsection{Main lobe to side lobe ratio}
The third criterion of main lobe to side lobe ratio (MSR) is displayed in figure \ref{fig:dynamics_MeanFD}. It is noticeable that in the frequency range $500\,$Hz $<f<1000\,$Hz the MSR is increased by up to $\Delta MSR = 4.5\,$dB by using the BEM steering vectors in the formulation III compared to the free-field steering vectors. The effect in this frequency range is dominated by the low MSR for beamforming maps using the free-field steering vectors, with a minimum of only $2\,$dB at $f = 660\,$Hz. In the corresponding beamforming map for source position 1 (fig. \ref{fig:S1_660Hz_FD_FF_steering_Conventional}) it can be seen that several side lobes at high levels are found. For this source position the MSR is even negative because the level at the actual source position is lower than in the lobes which are interpreted as side lobes. Only by averaging over four source positions, the MSR in figure \ref{fig:dynamics_MeanFD} remains in the positive range.

Above $f = 1000\,$Hz, the MSR of the beamforming measurements using the BEM simulated steering vector drops below that of the free-field cases. This behavior is illustrated in the beamforming map at $f = 1800\,$Hz (fig. \ref{fig:S4_1800Hz_FD_FF_steering_Conventional} and \ref{fig:S4_1800Hz_FD_BEM_steering_Conventional}) for the free field and BEM- simulated Green's function case, respecivaly. In the free-field case, the sidelobes are arranged more or less concentric around the main lobe. In the BEM case this pattern is not visible. Instead, some side lobes are found, of which a few have a higher level than the side lobes in the free-field case.

For the synthetically generated CSMs the results calculated with BEM steering vectors remain at a high level between $MSR = 6\,$dB and $MSR = 8.8\,$dB.

\subsubsection{Source to pattern ratio}
The last criterion is the source to pattern ratio (SPR, eq. \ref{eq:SPR}), which describes the logarithmic ratio between the sound pressure at the actual source position and the average sound pressure at all other focus points (within the heat exchanger box). This criterion is intended to take into account that not only the highest side lobe but any contribution to the beamforming map outside the source position (also within the main lobe) has a negative effect on the sound source localization. Furthermore, it summarizes the other criteria: The SPR is especially high when the highest sound pressure is found at the source position, the main lobe around it is narrow and the highest side lobe is low.

\begin{figure}
	\centering
	\begin{subfigure}[b]{0.45\textwidth}
		\centering
		\includegraphics[width=\textwidth]{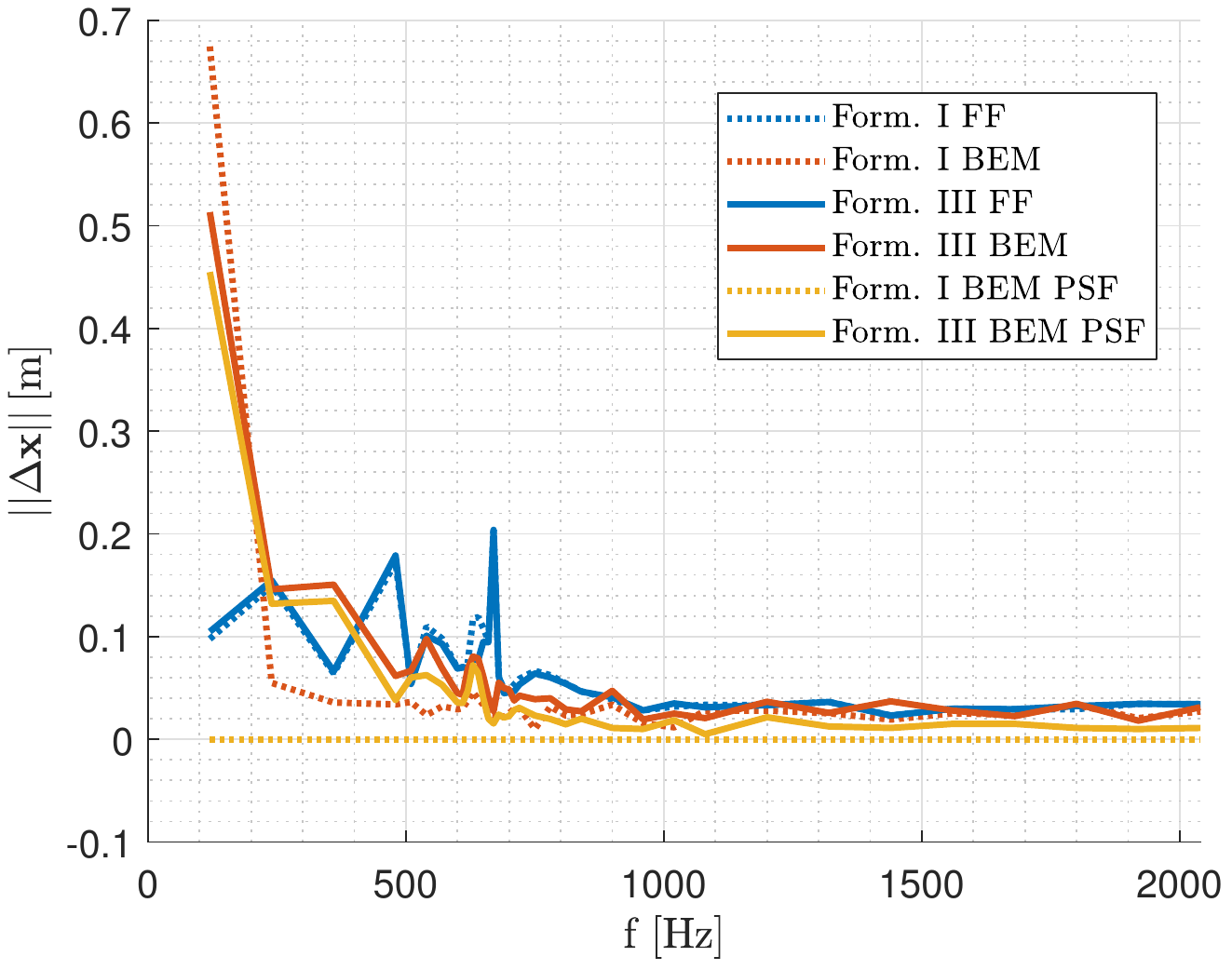}
		\caption{}
		\label{fig:dx_MeanFD}
	\end{subfigure}
	\begin{subfigure}[b]{0.45\textwidth}
		\centering
		\includegraphics[width=\textwidth]{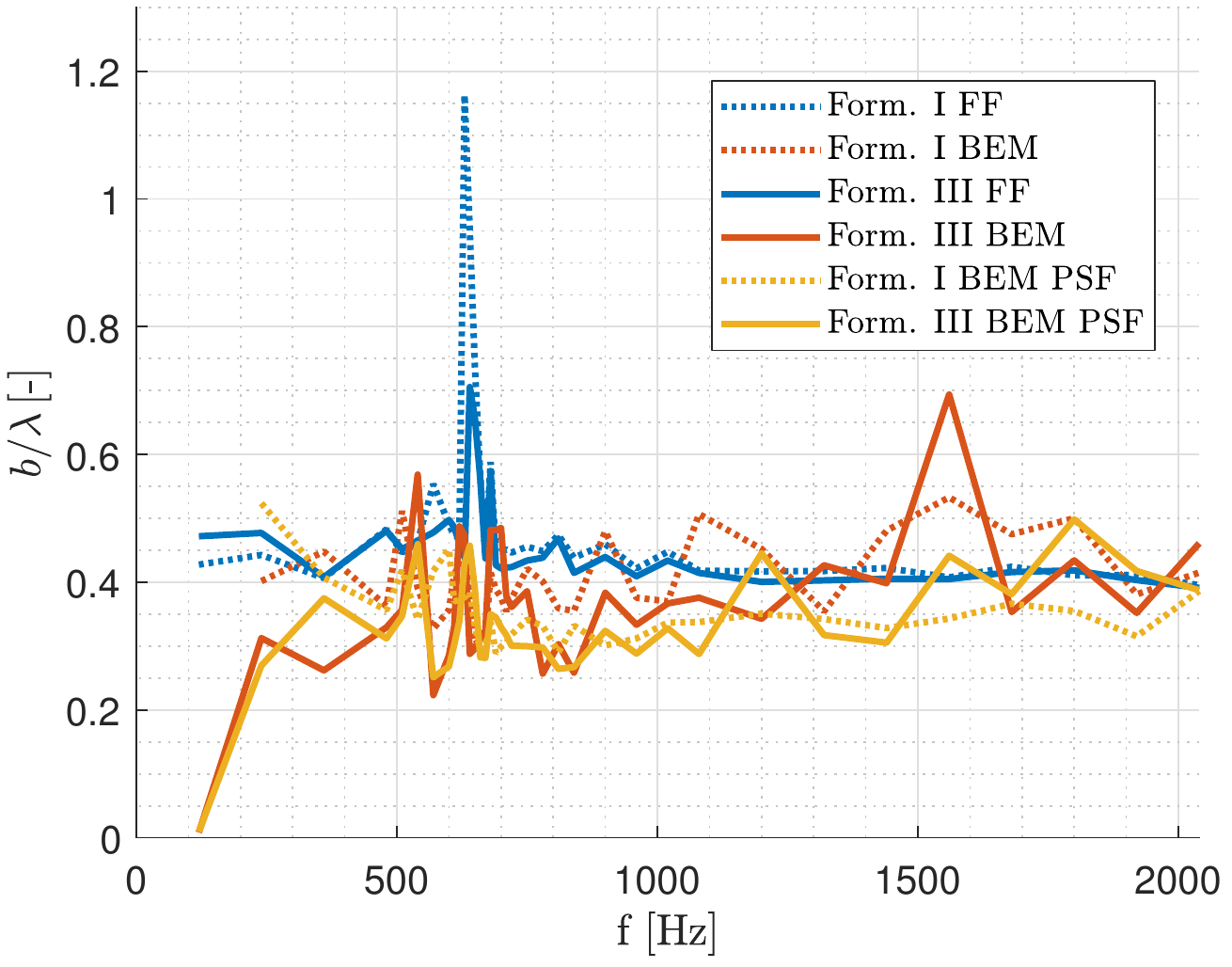}
		\caption{}
		\label{fig:resolutionHhMax_MeanFD}
	\end{subfigure}
	\hfill
	\begin{subfigure}[b]{0.45\textwidth}
		\centering
		\includegraphics[width=\textwidth]{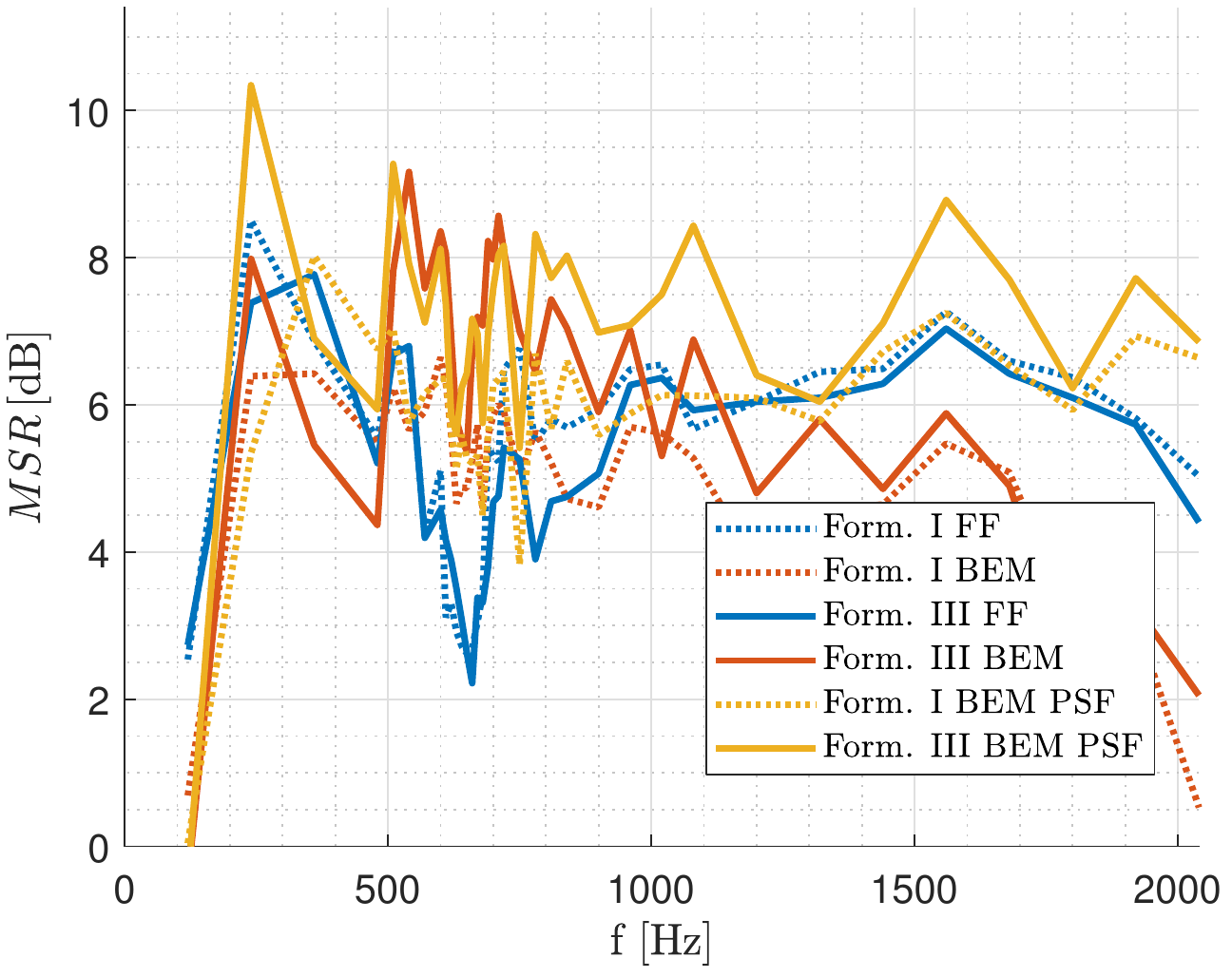}
		\caption{}
		\label{fig:dynamics_MeanFD}
	\end{subfigure}
	\begin{subfigure}[b]{0.45\textwidth}
		\centering
		\includegraphics[width=\textwidth]{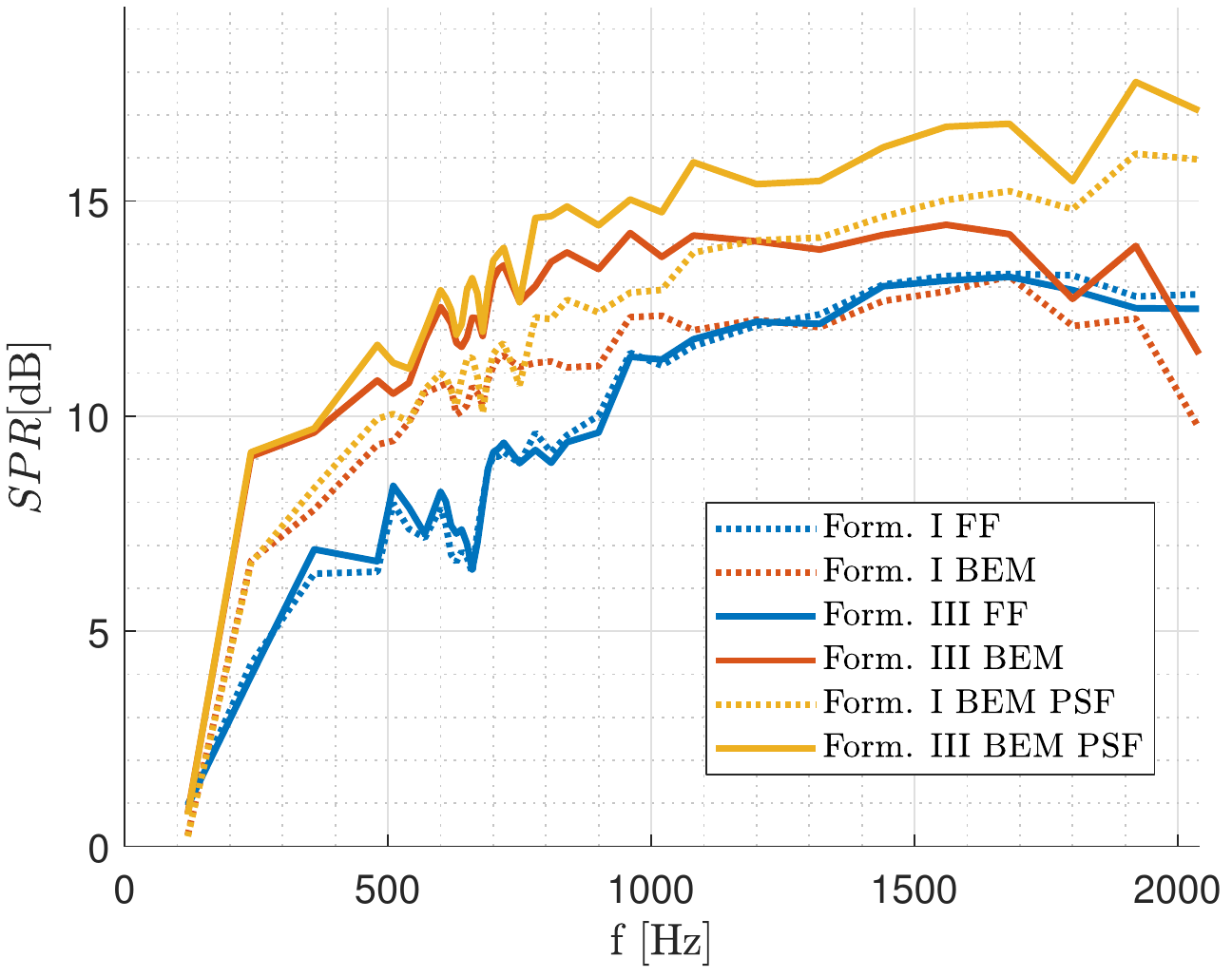}
		\caption{}
		\label{fig:SPRdB_MeanFD}
	\end{subfigure}
	\caption{Evaluation of measured data. Validation criteria ( (a): spatial deviation, (b): resolution, (c):main lobe to side lobe ratio and (d):source to pattern ration) over frequency for free-field (FF) and BEM- calculated (BEM) Green's functions and steering vector formulations I and III. Mean values over four source positions. Results from PSFs for BEM steering vectors given additionally for comparison.}
	\label{fig:FD}
\end{figure}

The curves are shown in figure \ref{fig:SPRdB_MeanFD}. The SPR is significantly increased for the investigated case if the steering vectors are calculated by using the BEM-simulated Greeen's function. For the steering vector formulation III, this is the case for almost the entire frequency range under consideration. Only at $f = 1800\,$Hz and $f = 2040\,$Hz the value of the free-field case is slightly higher. For the steering vector formulation I the SPR results are improved below $f = 1200\,$Hz by using the BEM-simulated compared to the corresponding free field case. 

\begin{figure}
	\centering
	\begin{subfigure}[b]{0.25\textwidth}
		\centering
		\includegraphics[width=\textwidth]{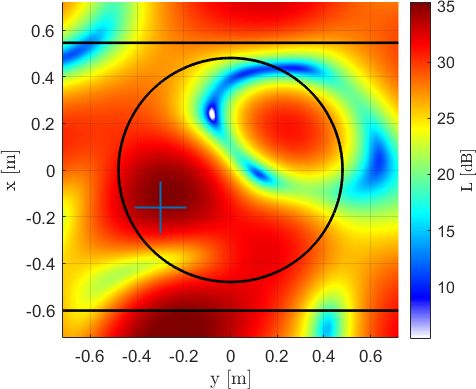}
		\caption{}
		\label{fig:S2_570Hz_FD_FF_steering_Conventional}
	\end{subfigure}
	\begin{subfigure}[b]{0.25\textwidth}
		\centering
		\includegraphics[width=\textwidth]{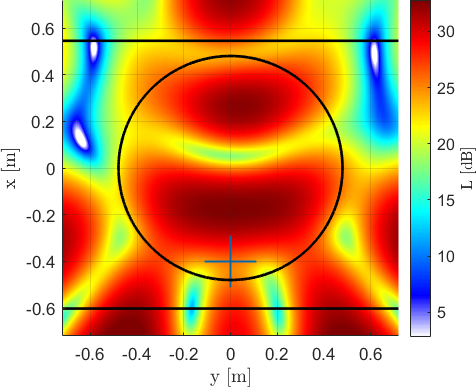}
		\caption{}
		\label{fig:S1_660Hz_FD_FF_steering_Conventional}
	\end{subfigure}
	\begin{subfigure}[b]{0.25\textwidth}
		\centering
		\includegraphics[width=\textwidth]{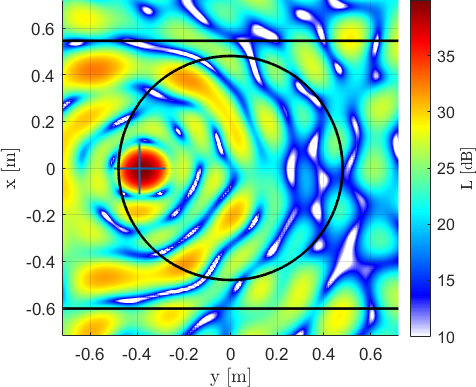}
		\caption{}
		\label{fig:S4_1800Hz_FD_FF_steering_Conventional}
	\end{subfigure}
	\hfill
	\begin{subfigure}[b]{0.25\textwidth}
		\centering
		\includegraphics[width=\textwidth]{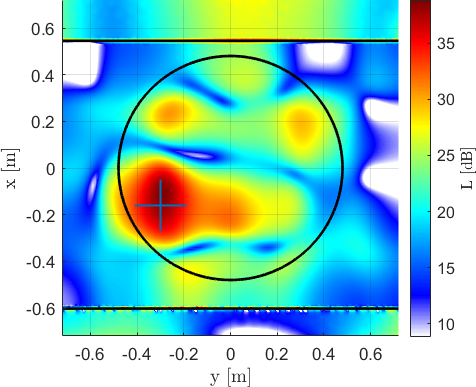}
		\caption{}
		\label{fig:S2_570Hz_FD_BEM_steering_Conventional}
	\end{subfigure}
	\begin{subfigure}[b]{0.25\textwidth}
		\centering
		\includegraphics[width=\textwidth]{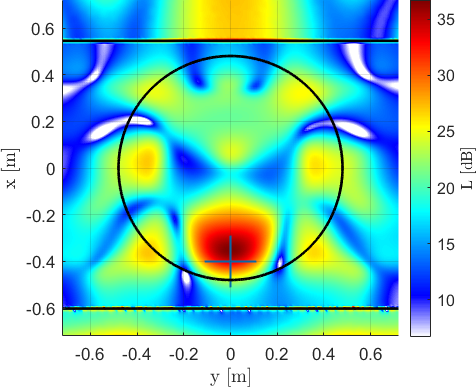}
		\caption{}
		\label{fig:S1_660Hz_FD_BEM_steering_Conventional}
	\end{subfigure}
	\begin{subfigure}[b]{0.25\textwidth}
		\centering
		\includegraphics[width=\textwidth]{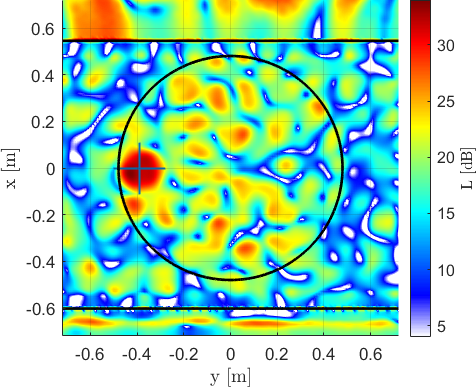}
		\caption{}
		\label{fig:S4_1800Hz_FD_BEM_steering_Conventional}
	\end{subfigure}
	\caption{Sample beamforming maps at different source positions with steering vector formulation III. Comparison between free-field (upper part) and BEM computed (lower part) Green's functions. (a)/(d): $f = 570\,$Hz, (b)/(e): $f = 660\,$Hz, (c)/(f): $f = 1800\,$Hz.}
	\label{fig:Maps_FD}
\end{figure}

The improvement can be illustrated or verified by looking again at the beamforming maps calculated with steering vector formulation III using the measurements and free field or BEM-simulated Greens's functions in figure \ref{fig:Maps_FD}, respectively. At $f = 570\,$Hz the SPR improves by $\Delta SPR = 4.5\,$dB and for $f = 660\,$Hz by $\Delta SPR = 5.9\,$dB. First of all, it is noticeable that for the BEM simulated steering vectors case the beamforming output outside the source position is significantly lower than for the free-field case. Additionally, in these cases the source positions are found more accurately.

Furthermore, it is noticeable from figure \ref{fig:SPRdB_MeanFD} that, when using BEM-simulated Green's functions, the formulation of the steering vectors has a significant influence on the results. The deviation between the two steering vector formulation  is up to $\Delta SPR = 2.7\,$dB. When using the  free-field steering vectors, the difference between the two steering vector formulations is always below $\Delta SPR = 0.6\,$dB.

In addition, it is remarkable that up to $f = 750\,$Hz the beamforming results using steering vectors based on the BEM-simulated Green's functions correspond well between measured and simulated microphone signals (CSMs). Above this range, the SPRs start to differ significantly: While SPR of the beamforming simulation still increases to higher frequencies, remains the SPR of the beamforming measurements more on the same level and start to decrease above $f=1800\,$Hz.

\subsubsection{Conclusions from beamforming measurement results with steering vector formulations I and III}
\label{subsubsec:Meas_res}
The validation criteria tested on the beamforming maps showed, that also for the measured microphone data the beamforming results can be significantly improved by using the BEM simulated tailored Green's functions. It was visible that some criteria show the results more transparently than others. The results of the various validation criteria indicate that the improvement is especially achieved in the low frequency range $f < 1000\,$Hz. Looking at the individual beamforming maps gives an indication for a reason: multiple lobes indicating different sound source appear in low frequency beamforming maps when using the free field Green's function. This can be explained as follows: The sound emitted by the loudspeaker is reflected by the walls of the heat exchanger box. Therefore, the sound waves reach the the microphones not only on the direct path but also indirectly, via reflections. The reflected sound waves hit the array from different directions than the direct sound waves (leading to different phase differences at the array microphones) and are therefore interpreted by free-field steering vectors as actual sources. However, in the BEM-simulated Green's function, the reflected wavefronts are taken into account, too. Or from a different point of view, the array can be interpreted as virtually mirrored on the heat exchanger box walls, thus enlarging the aperture. This improves the resolution of the beamforming. It is noticeable that the closer the source is to the wall perpendicular to a certain coordinate direction, the better the resolution for that coordinate direction is improved. A possible explanation for this is that in such case the distance to the opposite wall is large and thus the propagation time difference between direct sound and reflected wave becomes large. This offers the beamformer more different information. This increases the virtual mirrored array.

Another possible reason for the improved resolution is related to the geometric shadowing of single microphones in the large array ring: BEM Green's functions model the diffraction of the sound at the edges of the box, so that these microphone signals match the correct propagation model in the beamforming process and the beamforming results are improved due to the larger aperture and increased number of useful microphone signals.

From the fact that the results at low frequencies are significantly improved, it can be concluded that the simulation of the Green's functions in this range match the reality with sufficient accuracy. At higher frequencies, the use of simulated steering vectors hardly improves or, in the case of MSR, even worsens the beamforming results. This can have several reasons:

On the one hand, the deviation between the simulated and the real Green's functions could be too large. As the frequency increases, the wavelength decreases and thus the phase deviation increases for a fixed geometric error. Geometrical errors in the alignment of the microphone array, the heat exchanger box or the loudspeaker therefore have a greater effect. Because the BEM-simulated Green's functions include reflection and diffraction, they are much more sensitive to geometric inaccuracies. The result are most probably further degraded because the reflections are over-predicted. In the current BEM-simulations, total reflection is assumed, which most probably does not correspond to real behavior. It can also be assumed that with increasing frequency the directivity of the loudspeaker increasingly differs from a monopole characteristic. This all sums up to a difference between real wave propagation and the propagation model used in the beamforming algorithm, even when using the BEM-simulated Green's functions. 

On the other hand, beamforming with free-field Green's functions is known to be robust at high frequencies ($f>1000\,$Hz) which can also be seen from the fact that the corresponding validation criteria do not show any outliers in this range. It can therefore be assumed that the free-field beamforming is not much disturbed by the geometrical setup in this frequency range. One possible explanation is that the diffraction on the edges of the heat exchanger duct decreases with decreasing wavelength.

the phase shift of the reflected sound waves at the individual microphones is sufficiently different from that of the first wavefront. In this case, the BEM-based method cannot improve the results if the reflections cannot be predicted with high precision.

\subsection{Measurement results for steering vector formulation II}
\label{subsec:results_form_II}
\begin{figure}
	\centering
	\begin{subfigure}[b]{0.25\textwidth}
		\centering
		\includegraphics[width=\textwidth]{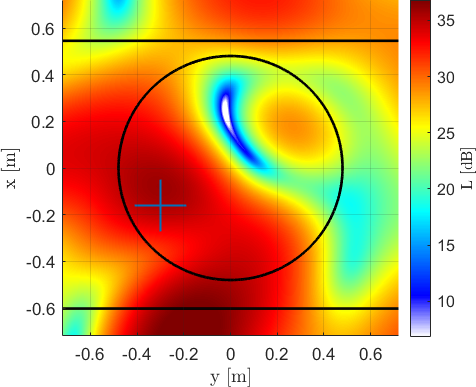}
		\caption{}
		\label{fig:S2_570Hz_FD_FF_steering_DelaySum}
	\end{subfigure}
	\begin{subfigure}[b]{0.25\textwidth}
		\centering
		\includegraphics[width=\textwidth]{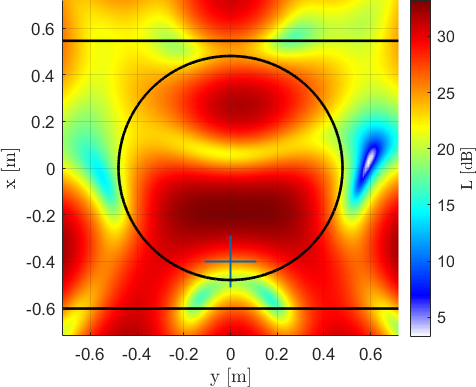}
		\caption{}
		\label{fig:S1_660Hz_FD_FF_steering_DelaySum}
	\end{subfigure}
	\begin{subfigure}[b]{0.25\textwidth}
		\centering
		\includegraphics[width=\textwidth]{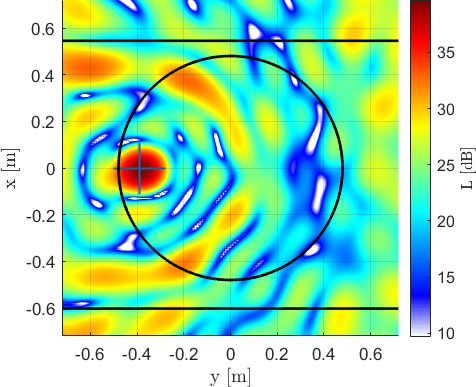}
		\caption{}
		\label{fig:S4_1800Hz_FD_FF_steering_DelaySum}
	\end{subfigure}
	\hfill
	\begin{subfigure}[b]{0.25\textwidth}
		\centering
		\includegraphics[width=\textwidth]{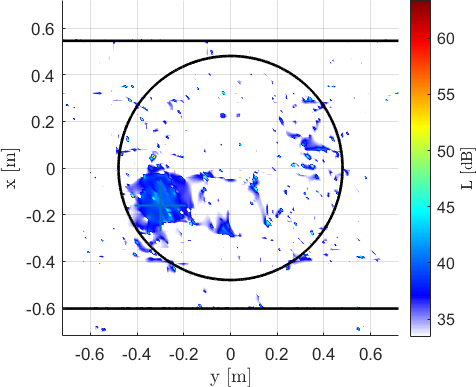}
		\caption{}
		\label{fig:S2_570Hz_FD_BEM_steering_DelaySum}
	\end{subfigure}
	\begin{subfigure}[b]{0.25\textwidth}
		\centering
		\includegraphics[width=\textwidth]{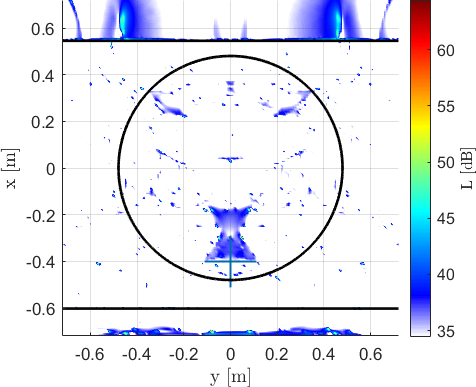}
		\caption{}
		\label{fig:S1_660Hz_FD_BEM_steering_DelaySum}
	\end{subfigure}
	\begin{subfigure}[b]{0.25\textwidth}
		\centering
		\includegraphics[width=\textwidth]{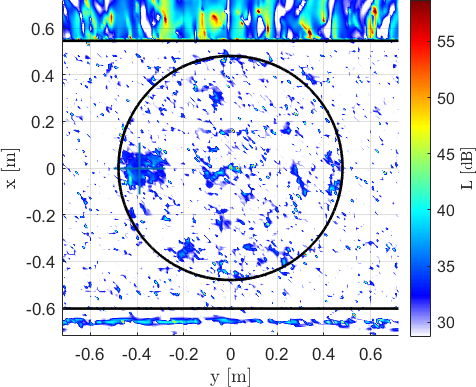}
		\caption{}
		\label{fig:S4_1800Hz_FD_BEM_steering_DelaySum}
	\end{subfigure}
	\caption{Sample beamforming maps at different source positions with steering vector formulation II. Comparison between free-field (upper part) and BEM computed (lower part) Green's functions. (a)/(d): $f = 570\,$Hz, (b)/(e): $f = 660\,$Hz, (c)/(f): $f = 1800\,$Hz.}
	\label{fig:Maps_FD_II}
\end{figure}

After the results for the steering vector formulations I and III have been shown, results for formulation II (eq. \ref{eq:SarradjII}) will be presented. 
Figure \ref{fig:Maps_FD_II} shows the corresponding beamforming maps for formulation II using free-field and BEM-simulated Green's functions, respectively. Source positions and frequencies of the presented beamforming maps are analogous to the results of formulation III (fig. \ref{fig:Maps_FD}).

The beamforming maps of the free-field Green's functions (upper line) are comparable to those of formulation III in figure \ref{fig:Maps_FD}. Contrary to that,  the results based on tailored Green's functions from BEM simulations (lower line) appear to differ greatly from the previous results. First of all, at some focus points, sound power levels of $L_{\text{max}} > 60 \,$dB are indicated. Comparing to that the highest sound power levels in the beamforming maps for formulation III (figure \ref{fig:S1_660Hz_FD_BEM_steering_Conventional}) are at $L_{\text{max}} \approx 35\,$dB. 

 A closer look reveals that the deviation of the levels found at the actual source position between formulation II and III is $\Delta L < 2\,$dB for all results from BEM- simulated Green's functions. However, the actual source position is difficult to identify from figure \ref{fig:S1_660Hz_FD_BEM_steering_DelaySum} because several other sound sources are indicated in the beamforming map  at similar levels  or even higher.

It can be concluded that formulation II, while revealing reasonable results when using the the free-field Green's functions, is not useful for the shown case when using the BEM-simulated Green's functions. 

This results may explained as follows. The main idea of the steering vector formulation II is to  weight the microphones signals in the beamforming process according to their (distance related) amplitude reduction. Thus signals recorded further by microphones away from the source are amplified. However, these microphones are mostly affected by higher uncertainties due to shadowing and refraction, which leads apparently to the disturbed beamforming maps. 
Therefore, formulation II will not be further considered. The evaluation of the according beamforming validation criteria was omitted.

\subsection{Summary and discussion of the results}
\label{subsec:disc_FD}
In this paper, five different validation criteria are introduced as measures to compare different beamforming maps. These validation criteria are applied to beamforming maps that are set up using three different steering vector formulations (I-III). A generalised formulation of these steering vectors was used to prove the preservation of their properties regarding the source position and source strength when using arbitrary differentiable Green's functions. 

The Greens's functions are calculated for a heat exchanger test case using BEM simulations in the frequency domain. These BEM simulated - tailored - Green's functions are afterward used to calculate the steering vectors. For comparison free-field Green's functions are used to calculate the according standard steering vectors.  

First, the method was validated. The synthetic validation case uses the same BEM simulations to calculate the propagation to the microphones as well as to calculate the steering vectors for the beamforming calculations. Using the first and second validation criterion, it was illustrated that, depending on the steering vector formulation, the source position or strength is found when the exact, tailored Green's functions are used to calculate the corresponding steering vectors.

Microphone array measurements were performed using a setup similar to the BEM test case, with a circular microphone array and a speaker as sound source at four different positions within the heat exchanger duct. 

For the different steering vector formulations, beamforming maps are calculated for the four different source positions in the frequency range up to $f=2040\,Hz$. The beamforming maps for formulation II of BEM simulated Green's functions showed poor results. Therefore, the values of four validation criteria are calculated for formulation I and III only and averaged for each frequency, respectively. These beamforming results, using the BEM- simulated steering vectors, are compared to the corresponding beamforming results with steering vectors based on the free field Green's functions, respectively. 

Based on the validation criteria it could be shown that the use of tailored Green's functions  improves the beamforming maps especially in the lower frequency range. In the mid to high frequency range above $1-2\,$kHz the use of the tailored Green's functions did not show advantages compared to use of free-field Green's functions.

Two reason were discussed: First, free-field beamforming is known to be robust at higher frequencies and the beamforming maps often provide already high resolution and sufficient dynamic and source to pattern ratio. On the other hand, at higher frequencies the beamforming using tailored Green's functions appears to be sensitive to deviations between the BEM simulated Green's functions and the real sound propagation. This argument is supported by the evaluations of the PSF where both the Green's functions and the sound propagation to the microphones are based on the same BEM simulation. Evaluating these beamfoming maps (PSFs) and evaluating the according validation criteria it could be shown that, even at higher frequencies, results could be improved by using the BEM simulated Green's functions. 

For practical reasons it would be the aim of the presented method to determine the Green's functions for microphone array measurements in a fan test bench by simulations only. From the shown results it could be concluded that in the presented case the use of BEM simulated Greens' functions for beamforming does not require calibration with measurement data. 

Nevertheless, it can be assumed that a calibration of the simulation results with measured Green's functions on the basis of some focus point-microphone combinations could improve the measurement results also at higher frequencies. However, this would increase the effort to perform the measurement. 

\section{Conclusion}
\label{sec:conclusion}
In the present work a method for sound source localization under disturbed propagation conditions has been presented. The test case was a fan test bench where a generic source is placed inside a short heat exchanger duct and microphone array measurements were performed outside the duct in a semi anechoic chamber. 

The Green's functions used in this paper were simulated using BEM calculations, considering all reflection, diffraction, and shadowing effects. The numerically determined Green's functions are applied to beamforming on an experimental dataset similar to the numerically simulated. No calibration was performed to tweak the simulated Greeen's functions using measurement data. Using a generalized steering vector formulation, it was shown that the properties of a correct source location and amplitude are valid for arbitrary differentiable Green's functions. 

In the process of validating the resulting maps, several validation criteria were set up for assessment of the effect of the simulated Green's functions versus a free-field monopole assumption.
It was shown that for the selected test case the results in the lower frequency range can be significantly improved compared to those of free-field beamforming by using numerically calculated Greens functions. 

\section{Appendix} 
\subsection{Steering vector}
\label{section:appendix_1}

The beamforming response $A$ at location $\mathbf{y} \in \mathbb{R}^3$ with a single source with unit amplitude at position $\mathbf{y}_s \in \mathbb{R}^3$ is
\begin{equation}\label{eqn:beamforming_single_summation}
    A(\mathbf{y}_s,\mathbf{y}) = \left|\mathbf{w}^*(\mathbf{y}) \mathbf{g}(\mathbf{y}_s)\right|^2 = \left| \sum_{m=1}^M w_m^*(\mathbf{y})g_m(\mathbf{y}_s)\right|^2
\end{equation}
where $g_m(\mathbf{y}) = |g_m(\mathbf{y})|e^{j\phi_m(\mathbf{y})}$ is the Green's function from the source position $\mathbf{y}$ to the $m$-th microphone in polar form. In general, the steering vector $w_m(\mathbf{y})$ can be defined in polar form
\begin{equation}
    w_m(\mathbf{y}) = f_m(\mathbf{y}) e^{j\phi_m(\mathbf{y})}
\end{equation}
with amplitude scale functions $f_m:\mathbb{R}^3 \rightarrow \mathbb{R}^+$. The beamforming response is therefore
\begin{equation}
\begin{split}
    A(\mathbf{y},\mathbf{y}_s) &= \left| \sum_{m=1}^M w_m^*(\mathbf{y})g_m(\mathbf{y})\right|^2 = \left| \sum_{m=1}^M e^{j(\phi_m(\mathbf{y}_s)-\phi_m(\mathbf{y}))}f_m(\mathbf{y}) |g_m(\mathbf{y}_s)|\right|^2 \\
    &= \underbrace{\left( \sum_{m=1}^{M} f_m(\mathbf{y}) |g_m(\mathbf{y}_s)|\cos (\phi_m(\mathbf{y}_s)-\phi_m(\mathbf{y}))\right)^2}_{A_1(\mathbf{y},\mathbf{y}_s)} +    \underbrace{\left( \sum_{m=1}^{M} f_m(\mathbf{y}) |g_m(\mathbf{y}_s)|\sin (\phi_m(\mathbf{y}_s)-\phi_m(\mathbf{y}))\right)^2}_{A_2(\mathbf{y},\mathbf{y}_s)}
\end{split}
\end{equation}
which is only a reformulation of equation (\ref{eqn:beamforming_single_summation}). The requirement that the beamforming response $b$ has a local maximum at the correct source position $\mathbf{y}_s$ is
\begin{equation}
    \begin{split}
        \left. \frac{\partial A(\mathbf{y},\mathbf{y}_s)}{\partial y_i}\right|_{\mathbf{y}=\mathbf{y}_s} = 0\quad  \forall i=1,2,3
    \end{split}
\end{equation}
The spatial derivative of $A_2(\mathbf{y},\mathbf{y}_s)$ regarding $\mathbf{y}$ vanishes at $\mathbf{y}_s$:
\begin{equation}
     \left. \frac{\partial A_2(\mathbf{y},\mathbf{y}_s)}{\partial y_i} \right|_{\mathbf{y}=\mathbf{y}_s} = 2\left( \sum_{m=1}^{M} f_m(\mathbf{y}_s) |g_m(\mathbf{y}_s)|\underbrace{\sin (\phi_m(\mathbf{y}_s)-\phi_m(\mathbf{y}_s))}_{=0}\right)\cdot (\ldots) = 0\quad \forall  i=1,2,3
\end{equation}
so that
\begin{equation}\begin{split}
    \left. \frac{\partial A(\mathbf{y},\mathbf{y}_s)}{\partial y_i}\right|_{\mathbf{y}=\mathbf{y}_s} &= \left. \frac{\partial A_1(\mathbf{y},\mathbf{y}_s)}{\partial y_i}\right|_{\mathbf{y}=\mathbf{y}_s} \\
    &= 2\underbrace{\left( \sum_{m=1}^{M} f_m(\mathbf{y}_s) |g_m(\mathbf{y}_s)|\right)}_{>0}\left( \sum_{m=1}^{M}|g_m(\mathbf{y}_s)| \left.\frac{\partial f_m(\mathbf{y})}{\partial y_i} \right|_{\mathbf{y}=\mathbf{y}_s} + \underbrace{\sin (\phi_m(\mathbf{y}_s)-\phi_m(\mathbf{y}_s))}_{=0}\cdot (\ldots)\right)
    \end{split}
\end{equation}
The first term is always positive. Hence, the spatial derivative of the beamforming response $b$ vanishes if and only if
\begin{equation}\label{eqn:correct_location_scale_function}
\sum_{m=1}^{M}|g_m(\mathbf{y}_s)| \left.\frac{\partial f_m(\mathbf{y})}{\partial y_i} \right|_{\mathbf{y}=\mathbf{y}_s} = 0
\end{equation}
The only requirement to derive equation (\ref{eqn:correct_location_scale_function}) is that the amplitude $|g_m(\mathbf{y})|$ and phase $\phi_m(\mathbf{y})$ function of the Green's function are differentiable regarding the source location $\mathbf{y}$. There is no dependency on the source type (monopol, dipol, linesource, etc.) or the way this Green's function is actually derived (analytical formulation, FEM or BEM calculation, etc.). A possible definition for the amplitude scale functions $f_m$ is
\begin{equation}\label{eqn:scale_function_definition}
    f_m(\mathbf{y}) := \frac{|g_m(\mathbf{y})|^{\beta-1}} {\left(\sum_{n=1}^{M}|g_n(\mathbf{y})|^{\beta}\right)^{\alpha}M^{1-\alpha}}
\end{equation}
Inserting definition (\ref{eqn:scale_function_definition}) in condition (\ref{eqn:correct_location_scale_function}) results in a simplified requirement for a steering vector with a local maximum at the source location:
\begin{equation}\label{eqn:correct_location_alpha_beta}
    \alpha=1-\frac{1}{\beta}
\end{equation}
A similar equation can be derived for the requirement that the beamforming response $b$ is correct at the source location $\mathbf{y}_s$. 
\begin{equation}\label{eqn:correct_amplitude_scale_function}
    A(\mathbf{y}_s,\mathbf{y}_s) = A_1(\mathbf{y}_s,\mathbf{y}_s) = 1 \Rightarrow \sum_{m=1}^{M} |g_m(\mathbf{y}_s)|f_m(\mathbf{y}_s)=1
\end{equation}
Again, inserting definition (\ref{eqn:scale_function_definition}) in condition (\ref{eqn:correct_amplitude_scale_function}) results in the simplified requirement for a steering vector formulation with the correct amplitude at the source location:
\begin{equation}\label{eqn:correct_amplitude_alpha_beta}
    \alpha = 1, \quad \beta \in \mathbb{R}
\end{equation}
Both requirements (\ref{eqn:correct_location_alpha_beta}) and (\ref{eqn:correct_amplitude_scale_function}) are fulfilled at the same time only if $\beta \rightarrow \pm \infty$. Unfortunately, this leads to clearly increased sidelobes in the point spread function and amplifies possible noise in the measurement drastically. Therefore it does not matter in practical applications.
The four steering vector formulations from Sarradj \cite{sarradj2012three} can be described as special case of equation (\ref{eqn:scale_function_definition}) by the parameters $\alpha$ and $\beta$:
\begin{align}
    \alpha &= 0,& \quad \beta = 1 & & \Rightarrow & & f_m &= \frac{1}{M} & \Rightarrow & & w_m^{I} &= \frac{1}{M} \frac{g_m}{\vert g_m \vert} \\
    \alpha &= 1,&\quad \beta = 0 & & \Rightarrow  & & f_m &= \frac{1}{M} \frac{1}{\vert g_m \vert} & \Rightarrow & & w_m^{II} &= \frac{1}{M} \frac{g_m}{\vert g_m \vert^2}\\
    \alpha &= 1,&\quad \beta = 2 & & \Rightarrow & & f_m &= \frac{|g_m|}{\|\mathbf{g}\|_2^2} & \Rightarrow & & w_m^{III} &= \frac{g_m}{\|\mathbf{g}\|_2^2} \\
    \alpha &= 1/2,&\quad \beta = 1 & & \Rightarrow & & f_m &= \frac{|g_m|}{\sqrt{M}\|\mathbf{g}\|_2} & \Rightarrow & & w_m^{IV} &= \frac{g_m}{\sqrt{M}\|\mathbf{g}\|_2}
\end{align}

\subsection{PSF from BEM simulated Green's functions}\label{section:appendix_2}
\begin{figure}[h]
	\centering
	\begin{subfigure}[b]{0.25\textwidth}
		\centering
		\includegraphics[width=\textwidth]{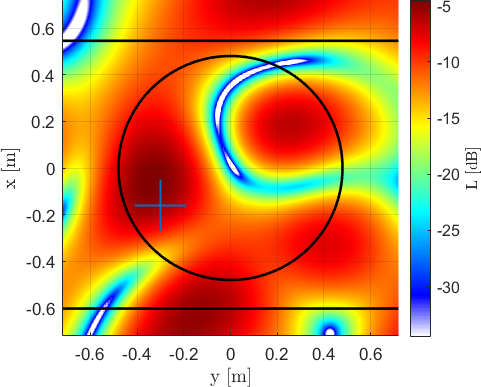}
		\caption{}
		\label{fig:S2_570Hz_PSF_BEM_FF_steering_Conventional}
	\end{subfigure}
	\begin{subfigure}[b]{0.25\textwidth}
		\centering
		\includegraphics[width=\textwidth]{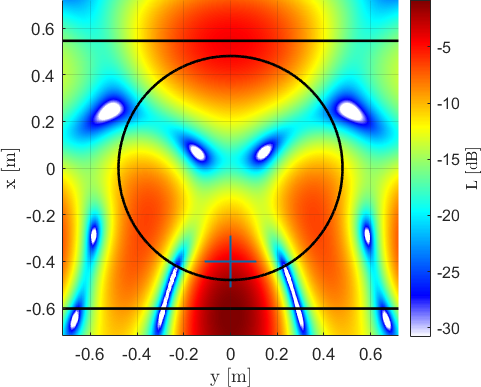}
		\caption{}
		\label{fig:S1_660Hz_PSF_BEM_FF_steering_Conventional}
	\end{subfigure}
	\begin{subfigure}[b]{0.25\textwidth}
		\centering
		\includegraphics[width=\textwidth]{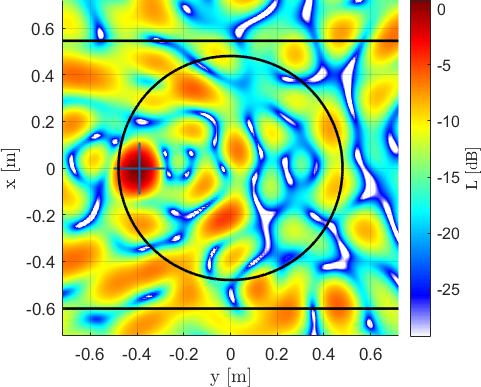}
		\caption{}
		\label{fig:S4_1800Hz_PSF_BEM_FF_steering_Conventional}
	\end{subfigure}
	\hfill
	\begin{subfigure}[b]{0.25\textwidth}
		\centering
		\includegraphics[width=\textwidth]{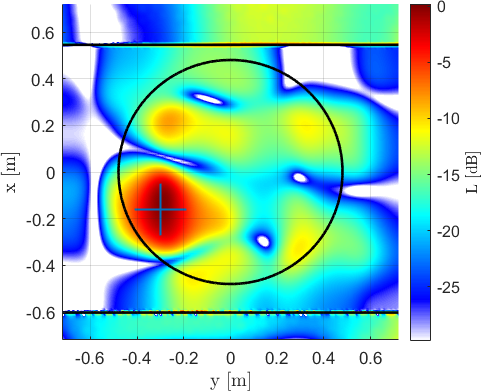}
		\caption{}
		\label{fig:S2_570Hz_PSF_BEM_BEM_steering_Conventional}
	\end{subfigure}
	\begin{subfigure}[b]{0.25\textwidth}
		\centering
		\includegraphics[width=\textwidth]{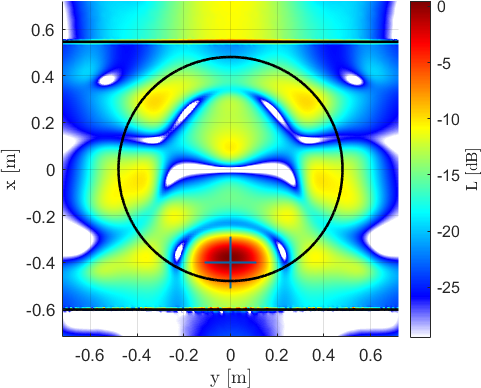}
		\caption{}
		\label{fig:S1_660Hz_PSF_BEM_BEM_steering_Conventional}
	\end{subfigure}
	\begin{subfigure}[b]{0.25\textwidth}
		\centering
		\includegraphics[width=\textwidth]{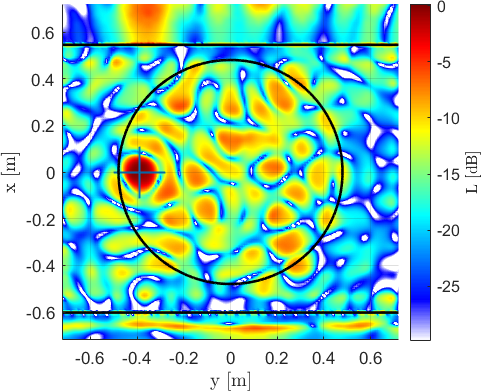}
		\caption{}
		\label{fig:S4_1800Hz_PSF_BEM_BEM_steering_Conventional}
	\end{subfigure}
	\caption{Sample point spread functions (PSF) from BEM simulated Greens functions at different source positions with steering vector formulation III. Comparison between free-field (upper part) and BEM simulated (lower part) Green's functions for steering vector calculation. (a)/(d): $f = 570\,$Hz, (b)/(e): $f = 660\,$Hz, (c)/(f): $f = 1800\,$Hz.}
	\label{fig:Maps_PSF}
\end{figure}

\bibliography{Literatur_Lehmann}

\end{document}